\documentclass[3p,10pt,review,fleqn]{elsarticle}

\graphicspath{{images/}}
\usepackage{caption}
\usepackage{subcaption}

\usepackage[modulo]{lineno}
\usepackage{graphicx}
\usepackage[usenames]{color}
\usepackage{amssymb}
\usepackage{amsmath}
\usepackage{epsfig}
\usepackage{empheq}
\usepackage{graphicx}
\usepackage{relsize}
\usepackage{amssymb}
\usepackage{amsthm}
\usepackage{amsmath}
\usepackage{amscd}
\usepackage{natbib}
\biboptions{sort&compress}
\usepackage{charter}
\usepackage{lineno,hyperref}
\modulolinenumbers[1]
\usepackage{array}
\journal{Structures}

\journal{Mechanics of Materials}

\begin{document}

\begin{frontmatter}

\title{The effective mechanical properties of solids with distributed rough cracks}
\author[ksh]{Kamal Shaker}
\address[ksh]{School of Civil Engineering, College of Engineering, University of Tehran, Tehran, Iran}
\author[hkh]{Hamed Khezrzadeh\corref{cor1}}
\address[hkh]{Faculty of Civil and Environmental Engineering, Tarbiat Modares University, Jalal Ale Ahmad Highway, P.O. Box 14115-143, Tehran, Iran}
\cortext[cor1]{Corresponding author.}
\ead{khezrzadeh@modares.ac.ir}

\begin{abstract}
A statistical multi-step self-consistent (SMSSC) micromechanical model for predicting elastic and plastic properties of a three-dimensional Representative Volume Element (RVE) is proposed in this research. A body with randomly distributed rough penny-shaped microcracks is considered for the first time in which the distribution of cracks is included through different well-known statistical distributions. Initially, the solution for a rough cohesive penny-shaped crack is developed, and the relationship between crack nominal length and roughness is derived. Consequently, Crack Opening Displacement (COD) and the corresponding volume crack opening are calculated as a function of surface roughness. Next, the SMSSC is used to study the effects of nominal crack length distribution on a fractured medium's mechanical properties with a known statistical distribution of crack sizes. This is done by determining the energy release and evaluating its share in material degradation (elastic, plastic). The elastic properties calculated using SMSSC indicated that the fractality of crack trajectories leads to lower values of volume crack opening and, consequently, resulted in a lower level of stiffness degradation compared with previous smooth crack approximations. It is also reached that the statistical distribution of rough cracks influences the mechanical properties due to the interrelation between length and roughness. Regarding the plastic behavior, the proposed micromechanical methodology is implemented to construct a capped pressure-sensitive yielding potential function that explicitly highlights the influence of crack size statistical distribution on the yielding of cracked solids. It is evident that the statistical distribution of crack sizes which are interrelated with crack roughness, can significantly alter the corresponding yield surface of fractured media.   
	
\end{abstract}

\end{frontmatter}

\section{Introduction}
Formulating analytical models that are capable of predicting the mechanical properties of materials has been a challenge. The presence of defects in solids' microstructure contributes to several issues such as stiffness reduction, strength degradation, and deterioration of the properties. Cracks, as one of the most critical types of defects, have attracted considerable attention among the scientific community, especially in the field of mechanics and geomechanics, for a long time (e.g., see \cite{fjar2008}- ch.6, and \cite{gueguen2011}). Following the introduction of the problem of cracked solids in the works of \cite{bristow1960,walsh1965a,walsh1965b}, different homogenization schemes have been proposed for a fractured media which significantly vary from each other that is pertaining to the intrinsic complexity of the problem raised from either geometry of a single crack, or cracks' orientation distribution and density.

Analytical and numerical methods have been applied in many studies to obtain the effective elastic properties of bodies containing different distribution of cracks and voids \cite{kovalenko1977,nemat1981,roberts2002}.
A vast majority of research on cracked solids has considered the effect of cracks as imposing excessive compliance and has tried to present a mean-field approximation of the apparent mechanical properties. The departure point for these works is the solution of a non-interaction or dilute approximation, and several schemes have been proposed to effectively address the interactions between cracks, including the self-consistent \cite{hill1965,OConnell1974,budiansky1976,hoenig1979}, differential \cite{salganik1973,hashin1988,zimmerman1990} and application of Mori-Tanaka \cite{mori1973} scheme for cracked solids \cite{benveniste1986}. Other major contributions assume a periodic setting for microcracks \cite{horii1990, nemat1993b, shen2001, berryman2002}. Kachanov introduced another direction by considering non-random distribution of microcracks \cite{kachanov1980, kachanov1992, kachanov1993} and using second rank crack density tensor. Modeling the effective properties of several types of microstructures were studied by Kashanov  and Sevostianov \cite{kachanov2005}. 
In the case of a flawed solid with a sparse distribution of cracks, the effective elastic modulus were calculated by Garbin and Knopoff \cite{garbin1973,garbin1975a,garbin1975b}.
The anisotropy induced by oriented cracks in the microstructure was studied in \cite{hudson1980,hudson1986,hudson1990}. Another source of anisotropy is loading, which was discussed in \cite{horii1983}. The concept of domain of microcrack growth (DMG) was developed in \cite{yu1995} to address the problem of a brittle solid weakened by microcracks.

The overwhelming majority of theoretical studies are devoted to the elasticity of solids with smooth penny-shaped cracks while fracture surfaces are usually irregular. However, in some research, the effect of irregularity is investigated through different methods, in \cite{mavko1978} the irregular cracks have been modelled by implementing a dislocation density function to generate the cross-section of crack and derived the expressions for several crack trajectories. 
The irregular cracks have been extensively studied in several other contributions (see \cite{gueguen2011} for a comprehensive review of different methods). Grechka et al. \cite{grechka2006} showed that even substantial deviations of geometries from their circular counterparts are still insignificant for overall elastic properties. These type of postulates is a direct consequence of applying the classical framework. At the same time, a rigorous mathematical representation of the rough cracks which are extensively developed in numerous research indicates that including roughness as an intrinsic characteristic of cracks imposes meaningful impact on the stress field. 

Introduction of fractal geometry by Mandelbrot \cite{mandelbrot1983}, and realization of that fracture surfaces in different materials are fractals presented in several pioneering works \cite{mandelbrot1984,saouma1994,saouma1990} were the first steps toward the so-called fractal fracture mechanics (see \cite{cherepanov1995} for a review). Mosolov \cite{mosolov1991} theoretically studied the change of the order of stress singularity at the rough crack tip. Balankin \cite{balankin1996a,balankin1996b,balankin1997} extended the scope of analysis to the case of self-affine fractal cracks. This theoretical framework was comprehensively extended in several later works on the subject \cite{yavari2000, yavari2002, yavari2002b,wnuk2003, wnuk2008,wnuk2009,yavari2010,khezrzadeh2011}. Phenomena such as initiation and propagation of cracks in compression \cite{yavari2000,yavari2002}, mirror-mist-hackle \cite{wnuk2008}, the existence of limiting roughness, and having speeds far lower than Rayleigh wave speed for limiting velocity of dynamic crack propagation \cite{yavari2010} can be explained in the setting of fractal fracture mechanics. The fractal geometry also was implemented in order to give an explanation of the observed size effects on tensile strength and fracture energy of heterogeneous quasi-brittle solids in \cite{carpinteri1994}. Recently, it is found in \cite{mecholsky2020} that the relation between fracture energy and branching energy can be established in terms of fractal dimension of crack surface.  

As it is known from the mathematical formulation of the linear elastic fracture mechanics (LEFM), singular fields will be reached that has no physical interpretation. Several models have been proposed to tackle this inconsistency with the physical reality; among them, the cohesive models \cite{barenblatt1962, dugdale1960, park2011} and gradient theories \cite{lurie2014, vasil2016}. In the cohesive models, a distributed cohesive force of finite length is applied at the crack's tip to exclude the stress singularity in this region. In addition, this model can be used in modeling the plastic zone in the elastoplastic fracture mechanics. Based on the generalized theory of elasticity, Vasil'ev and Lurie \cite{vasil2016} proposed a non-singular solution for the problem of an unbounded plate with a crack. In their solution, a structure parameter should be determined through experiments. Then, they calculated the strength of a brittle body with a crack and compared the results with the experiments \cite{vasil2020}. However, determining the structure parameter that exists in their study for a three-dimensional cracked body is of great complexity.

In addition to the elastic properties of defective media, their strength is also negatively correlated with the presence of defects. Cracks as a major type of defects have always been a subject of considerable research. Started by the seminal work of Griffith \cite{griffith1921}, this area experienced substantial progression within a century. Numerous experimental studies suggest that void growth is the micromechanism of fracture in ductile materials \cite{tvergaard1990, pardoen2000}. The Gurson model \cite{gurson1977} is a well-known statistical homogenization model, which was developed for modelling void growth at micro-scale. The model correlates the growth of voids with damage parameters to describe its influence on the stress carrying capacity of the material. Many modifications have been proposed to account for other processes of ductile fracture, among which one can mention the Gurson\textendash Tvergaard\textendash Needleman model \cite{tvergaard1984}. The GTN model is extended from the original form of the Gurson model by considering the void nucleation and the coalescence of the crack process. According to many investigations on most brittle, quasi-brittle and even some ductile materials, nucleation and coalescence of microcracks are responsible for material failure \cite{mcclintock1968,tvergaard1990,pardoen2000}. A damage theory for quasi-brittle solids containing randomly distributed penny-shaped cohesive cracks was proposed in \cite{li2004}. 

In this study, the effective properties of solids containing a randomly distributed cohesive rough penny-shaped microcracks of a known statistical distribution is addressed for the first time. This paper is structured as follows: In \S~\ref{sec2}, stress fields of an RVE containing a penny-shaped cohesive crack are reviewed, and a relation between the cohesive stress and the yield stress of the virgin material is established. Next, a relationship between crack length and its surface roughness is suggested. Using the solution for a fractal slit-like Barenblatt\textendash Dugdale crack, the stress and displacement fields of a fractal cohesive penny-shaped crack is obtained. In \S~\ref{sec3}, using a statistical multi-step self-consistent micromechanical approach, the elastic properties of the RVE containing a distribution of rough cracks are calculated. In \S~\ref{sec4}, to describe damage properties of cracked media, the effective yielding potential function for an RVE containing rough cohesive cracks with some definite length distributions is investigated. Finally, conclusions are given in \S~\ref{sec5}.

\section{Geometrical modeling of a cracked medium}
\label{sec1}
There are various types of inhomogeneities of different sizes in the microstructure of materials. The sizes of these inhomogeneities may be within several length scales, and in each length scale, they affect the mechanical properties differently. The size distribution of inhomogeneities, which may be continuous over a wide range of length scales, can be represented by some probability distribution functions (PDF). To avoid the complexity of manipulation with the continuous size distribution function, one can discretize them into some sub-intervals. The point is that dividing the whole interval into some sub-intervals should be so that the cracks considered in each sub-interval approximately impose a similar degree of degradation on the properties. 

In this paper, a specific type of inhomogeneity, named penny-shaped crack, is considered. There are cracks of different sizes, and every size of the crack has a specific effect on the deterioration of the material strength. It would be of great interest to study the effect of every length scale separately. Extensive research on the presentation of cracks in more accurate models has revealed that the trajectories of fracture surfaces are rough, and the roughness changes when the crack advances. This irregularity in the crack surface can be modeled by fractal geometry. Fractal geometry is a mathematical tool for modeling irregular objects at different length scales.

\begin{figure}[h]
	\begin{center}
		\includegraphics[width=0.8\linewidth]{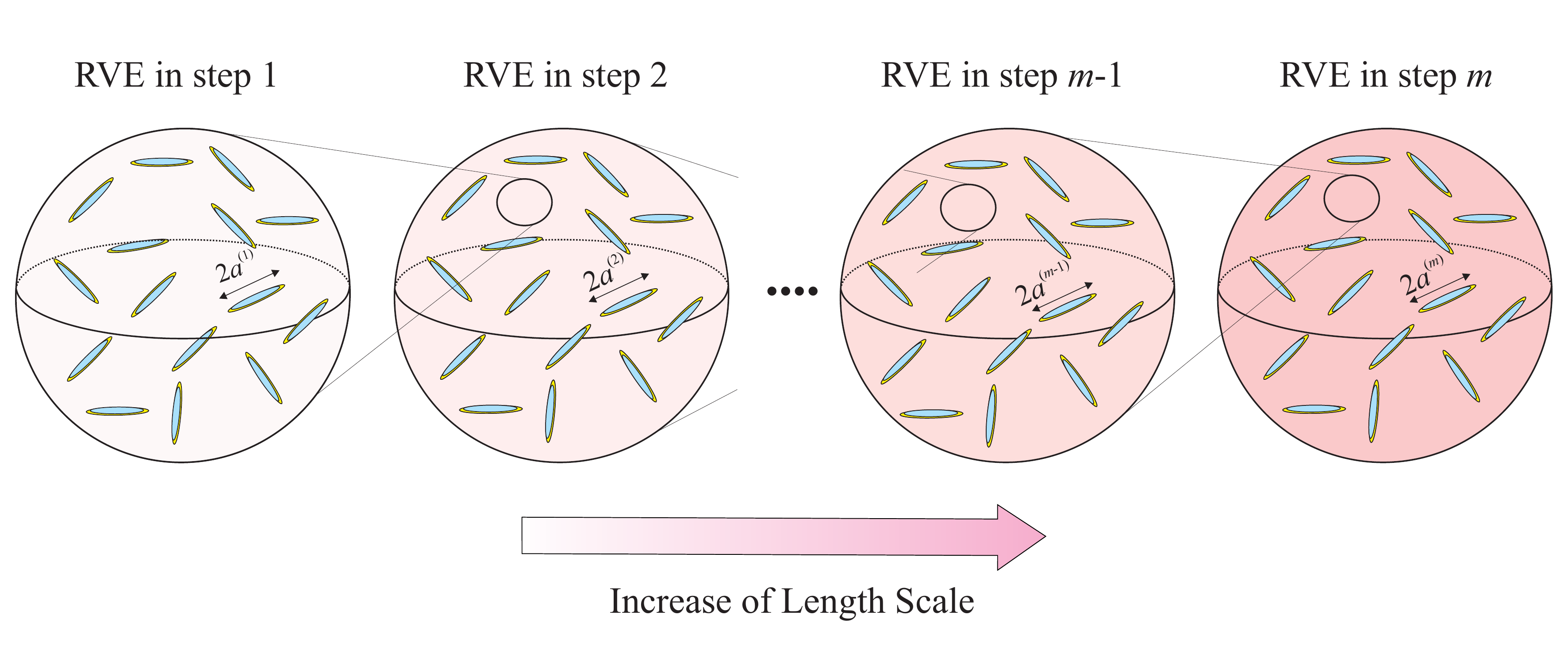}
		\caption{Hierarchy of RVEs: the RVE of the $i$-th step contains the microcracks belonging to the range $ [a^{(i)}-(a^{(m)}-a^{(1)})/m,a^{(i)}+(a^{(m)}-a^{(1)})/m] $. The crack radius at the middle of the interval is considered the representative crack radius for the whole homogenizing step.}
		\label{MS_RVE}
	\end{center}
\end{figure}
In this study, to construct a procedure for reaching the mechanical properties of a cracked solid, pre-defined upper and lower cutoffs for the crack sizes are considered. The crack size distribution of a material can be expressed by a proper probability distribution function. This crack size distribution function is discretized to some sub-intervals so that the microcracks located in each sub-interval approximately have a similar fractal dimension. To describe the geometrical model for the cracked body, consider the schematics shown in Fig.\ref{MS_RVE}. A hierarchical homogenization procedure is designed. Since lower and upper cutoffs for cracks sizes are denoted by, i.e. $a^{(1)}$ and $a^{(m)}$, respectively, $m$ sub-intervals are chosen for discretization. It should be noted that the procedure, depending on the volume fraction and size distribution of cracks, can be changed. An RVE, which is representative of that step, is introduced in each step. The RVE of the $i$-th step contains the microcracks that belong to that length scale. In each length scale, the matrix of the RVE is built of the material with effective mechanical properties obtained from the RVE of the previous step. The steps are continued until the whole length scales are involved. The effective material properties of the last step represent the homogenized properties of interest. Through this process, the effect of each length scale with assigned mechanical properties is separately implemented. 

To obtain the effective mechanical properties of the material by a self-consistent scheme, it is needed to determine the strain energy released by the microcracks in each length scale. This strain energy depends on the crack opening displacements of the cracks. Before that, it is necessary to obtain the displacement of the crack surface as a function of fractal parameters. In the following sections, the requirements and the mathematical framework of the homogenization method are provided.

\section{A penny-shaped cohesive crack under uniform triaxial tension}
\label{sec2}
An essential part of the homogenization procedure of a fractured media is the single penny-shaped cohesive crack solution. At the first part of this section, derivations for a smooth penny-shaped crack are given for a self-contained presentation. Next, solutions for a solid containing rough cracks will be derived.

\subsection{A smooth penny-shaped cohesive crack}
\label{subsec21}
\begin{figure}[h]
	\begin{center}
		\includegraphics[width=0.4\linewidth]{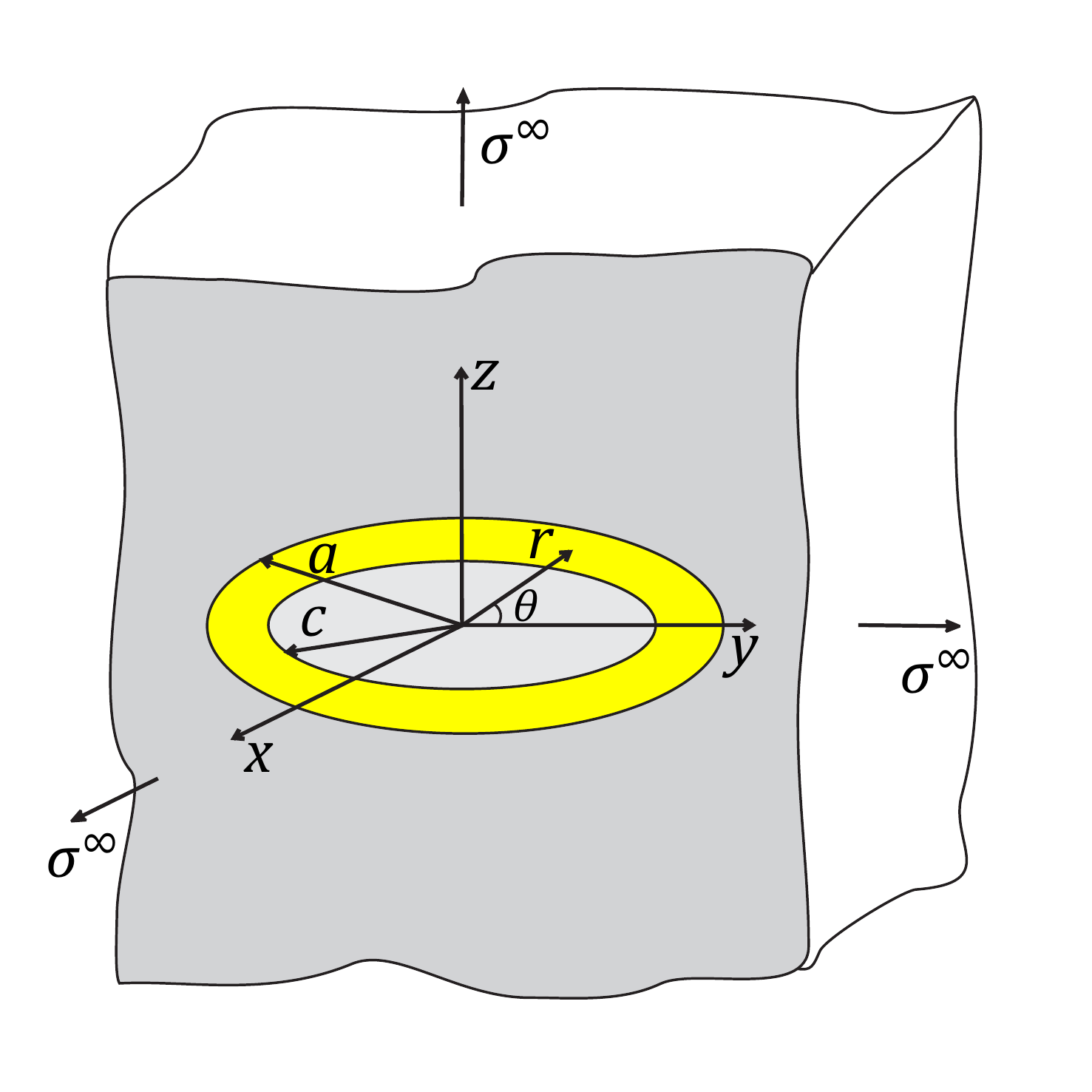}
		\caption{A penny-shaped cohesive crack in an RVE with cohesive ring width $ a-c $.}
		\label{P-SCr}
	\end{center}
\end{figure}
\begin{figure*}
	\begin{center}
		\includegraphics[width=0.8\linewidth]{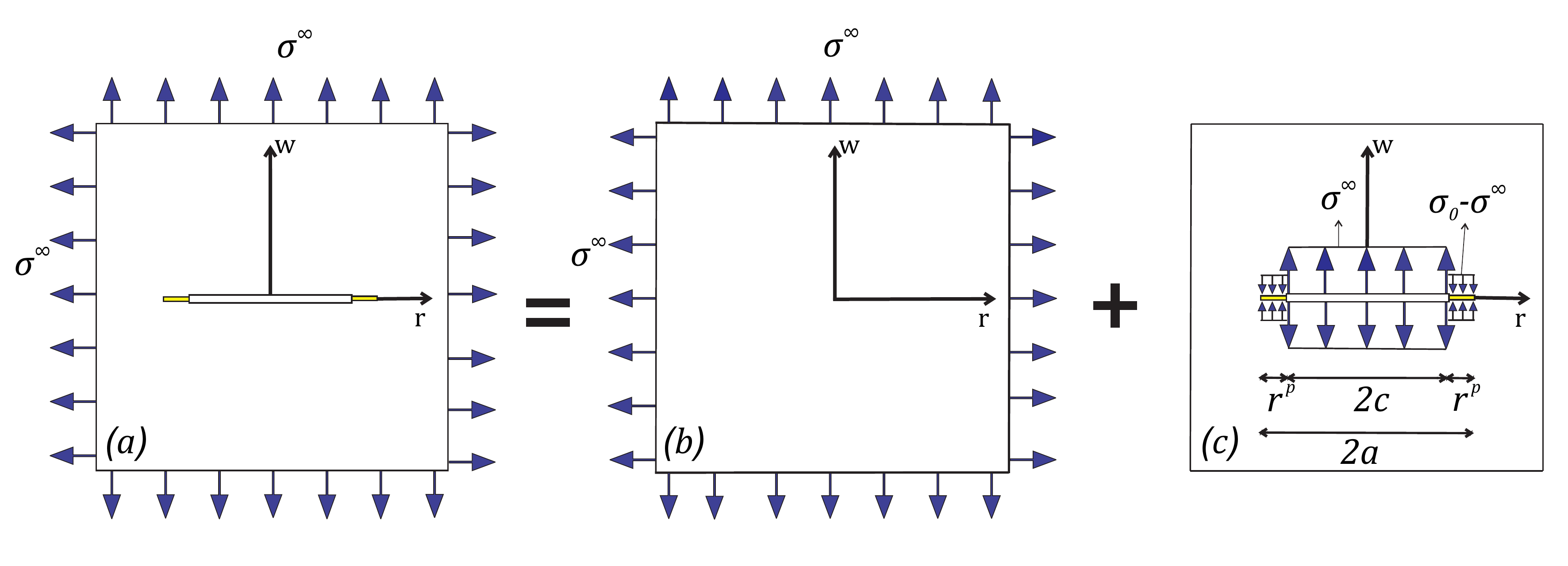}
		\caption{Penny-shaped crack decomposition: (b) Uniformly loading intact medium, and (c) Cohesive crack subjected to internal stresses}
		\label{ProDec}
	\end{center}
\end{figure*}

Consider a three-dimensional penny-shaped Barenblatt\textendash Dugdale crack of radius $ c $ (Fig.~\ref{P-SCr})
which is subjected to uniform far-field tensile stress, i.e., $ \sigma^\infty_{ij}=\sigma^\infty \delta_{ij} $.
Average theorem\footnote{The macro-stress tensor ($ \mathbf{\Sigma} $) is defined as volume average of micro-stress tensor in an RVE as
\begin{equation}
           \Sigma_{ij}=<\sigma_{ij}>=\dfrac{1}{V}\int_V \sigma_{ij} dV \nonumber .
\end{equation}
The macro-stress of the RVE is equal to the constant remote stress, i.e.,
\begin{equation}
        \Sigma_{ij}=<\sigma_{ij}>=\sigma_{ij}^{\infty} \nonumber.
\end{equation}
}
leads to $ \sigma^\infty=\Sigma_m $ where $  \Sigma_m $ is the macro-stress. There is a cohesive zone  with width $ r^p=a-c $ in an RVE around the crack, where $ a $ is the length of the extended crack. 
The problem can be solved by superimposing two problems, that are illustrated in Fig. \ref{ProDec}
The displacement and stress fields can be expressed in term of Popkovitch\textendash Neuber potential function (see \cite{green1968,kassir1975}).
Recalling the assumptions of Barenblatt\textendash Dugdale model, the stress singularities along the crack front, at $ w=0 $ and $ r=a $ must vanish. To have vanishing stress intensity factor $ K_I$, the following relationship should be hold between the applied stress ($ \sigma^{\infty} $), the yield zone stress ($ \sigma_0 $), and the physical and extended crack lengths (i.e., $c$ and $a$, respectively) \cite{chen1993},
\begin{equation}\label{cohesive size of penny}
          \frac{c}{a}=\sqrt{1-\frac{\sigma_{\infty}^2}{\sigma_0^2}}.
\end{equation}
The displacement $ u_w $ (in $ w=0 $ plane) is found as \cite{chen1993}
\begin{eqnarray}\label{displacement}
u_w(r)= \begin{cases}
\frac{2}{\pi}(\frac{1-\nu }{\mu})\left(\sigma^{\infty} \sqrt{a^2-r^2}-\sigma_0 \int_c^a \frac{\sqrt{t^2-c^2}}{\sqrt{t^2-r^2}} dt \right) \hspace{2.0cm}\text{for}\hspace{0.3cm} 0<r<c\\
\frac{2}{\pi}(\frac{1-\nu}{\mu})\left(\sigma^{\infty} \sqrt{a^2-r^2}-\sigma_0 \int_r^a \frac{\sqrt{t^2-c^2}}{\sqrt{t^2-r^2}} dt\right) \hspace{2.0cm}\text{for}\hspace{0.3cm} c<r<a,
\end{cases}
\end{eqnarray}
where $\mu$ and $\nu$ are shear modulus and Poisson's ratio, respectively.
As it is seen in fig. \ref{ProDec}, the problem is decomposed into two problems; so, the stresses in the main problem, named (a), are reached through superposition of the stress fields of two sub-problems (b) and (c). The stress components for the main problem can be reached as:
\begin{align}
	&\sigma_{ww}^{(a)}=\sigma_{ww}^{(b)}+\sigma_{ww}^{(c)}=\sigma_{0}\nonumber\\
	&\sigma_{rr}^{(a)}=\sigma_{rr}^{(b)}+\sigma_{rr}^{(c)}=\dfrac{1-2\nu}{2} \sigma^{\infty} + \sigma_{0}\nonumber\\
	&\sigma_{\theta \theta}^{(a)}=\sigma_{\theta \theta}^{(b)}+\sigma_{\theta \theta}^{(c)}=\dfrac{1-2\nu}{2} \sigma^{\infty} + 2\nu \sigma_{0}\nonumber\\
	&\sigma_{rw}^{(a)}=\sigma_{r\theta}^{(a)}=\sigma_{w\theta}^{(a)}=0.
\end{align}
Substituting these stress components in the well-known von-Mises criterion
\begin{equation}
    (\sigma_{ww}^{(a)}-\sigma_{rr}^{(a)})^2+(\sigma_{ww}^{(a)}-\sigma_{\theta \theta}^{(a)})^2+
    (\sigma_{\theta \theta}^{(a)}-\sigma_{rr}^{(a)})^2=2\sigma_Y^2
\end{equation}
leads to
\begin{equation}
4\left( \dfrac{\sigma_0}{\sigma^{\infty}}\right) ^2-2\left( \dfrac{\sigma_0}{\sigma^{\infty}}\right)+1-	\left( \dfrac{2}{1-2\nu} \dfrac{\sigma_Y}{\sigma^{\infty}}\right) ^{2}=0,
\end{equation}
where $ \sigma_Y $ is the yield stress of the virgin material in uniaxial tension. The positive root of this equation is
\begin{equation}
   \dfrac{\sigma_0}{\sigma^{\infty}}=\dfrac{1}{4} \left( 1+\sqrt{\left( \dfrac{4\sigma_Y}{(1-2\nu)\sigma^{\infty}}\right) ^2-3} \right).
\end{equation}
This gives the relation between the cohesive stress and the yield stress of the virgin material.
For different values of Poisson's ratio and considering $ 0<\dfrac{\sigma^{\infty}}{\sigma_0}<1 $, the cohesive stress  and the initial yield stress approximately have a linear relationship, i.e., $ \sigma_0\approx\dfrac{4}{\sqrt{12}(1-2\nu)}\sigma_Y $. Therefore, the concept of constant yield stress leads to a constant cohesive stress inside the cohesive zone \cite{li2004}.

\subsection{A fractal penny-shaped cohesive crack}
\label{subsec22}
There have been numerous research focusing on the removal of the physically unacceptable singularity of stress, and several methods were introduced. Among the significant contributions in the field one can mention plasticity correction \cite{orowan1954, irwin1958}, discrete nature of fracture phenomenon \cite{novozhilov1969}, the introduction of \textit{quantization procedure} \cite{pugno2004} and the consequent \textit{quantized fracture mechanics}.

Following the methodology developed in \cite{khezrzadeh2011}, an alternative approach is implemented by considering the cohesive zones at the crack tips while rest of the material remains elastic. The yielding cohesive stress of ($\sigma_0 $) is considered at the crack tips. As stated, many investigations on fractal cracks show that the order of stress singularity at the tip of fractal crack is different from that of a smooth crack \cite{mosolov1991, gol1991, balankin1997}.
The dominant term of the stress field around the tip of a fractal crack can be obtained through an asymptotic analysis. Although asymptotic analysis cannot give complete field solutions, it can be utilized to obtain the order of stress singularity $\alpha$ for fractal cracks. This order is found to depend on the degree of roughness \cite{yavari2000,yavari2002,yavari2010}. For the case of a self-affine fractal crack it is expressed in terms of Hurst exponent $H$ ($1/2\leq H\leq 1$).\footnote{In the case of rough surface cracks, the singularity is of order $\alpha$, i.e., $ r^{-\alpha} $, where it is $ r^{-1/2} $ in the LEFM. For self-similar and self-affine cracks, the fractal exponent $ \alpha $ is expressed as $ \alpha=\dfrac{2-D}{2}  $ ($ 1\leq D\leq 2 $) and $  \alpha=\dfrac{2H-1}{2H} $ ($ 1/2\leq H\leq 1 $), respectively. $D$ is the fractal dimension (see \cite{mandelbrot1985} for more details).}
For $ H=1 $ the fractal object possesses the properties of a 1-D Euclidean object while for $ H=1/2 $ the fractal set degenerates into a plane-filling 2-D object.
Many experimental studies \cite{bouchaud1990,maaloy1992,bouchaud1997,bouchaud2003,ponson2006} have indicated that a limiting roughness for the rough surfaces exists and it is about $H\simeq 0.8$, which is believed to be universal, i.e., independent of the fracture mode and the material.
This observation also found a theoretical justification when Yavari and Khezrzadeh \cite{yavari2010} used a branching argument and estimated a limiting roughness about $ H=0.8 $ for slit-like fractal cracks.

\begin{figure*}
	\begin{center}
		\includegraphics[width=0.8\linewidth]{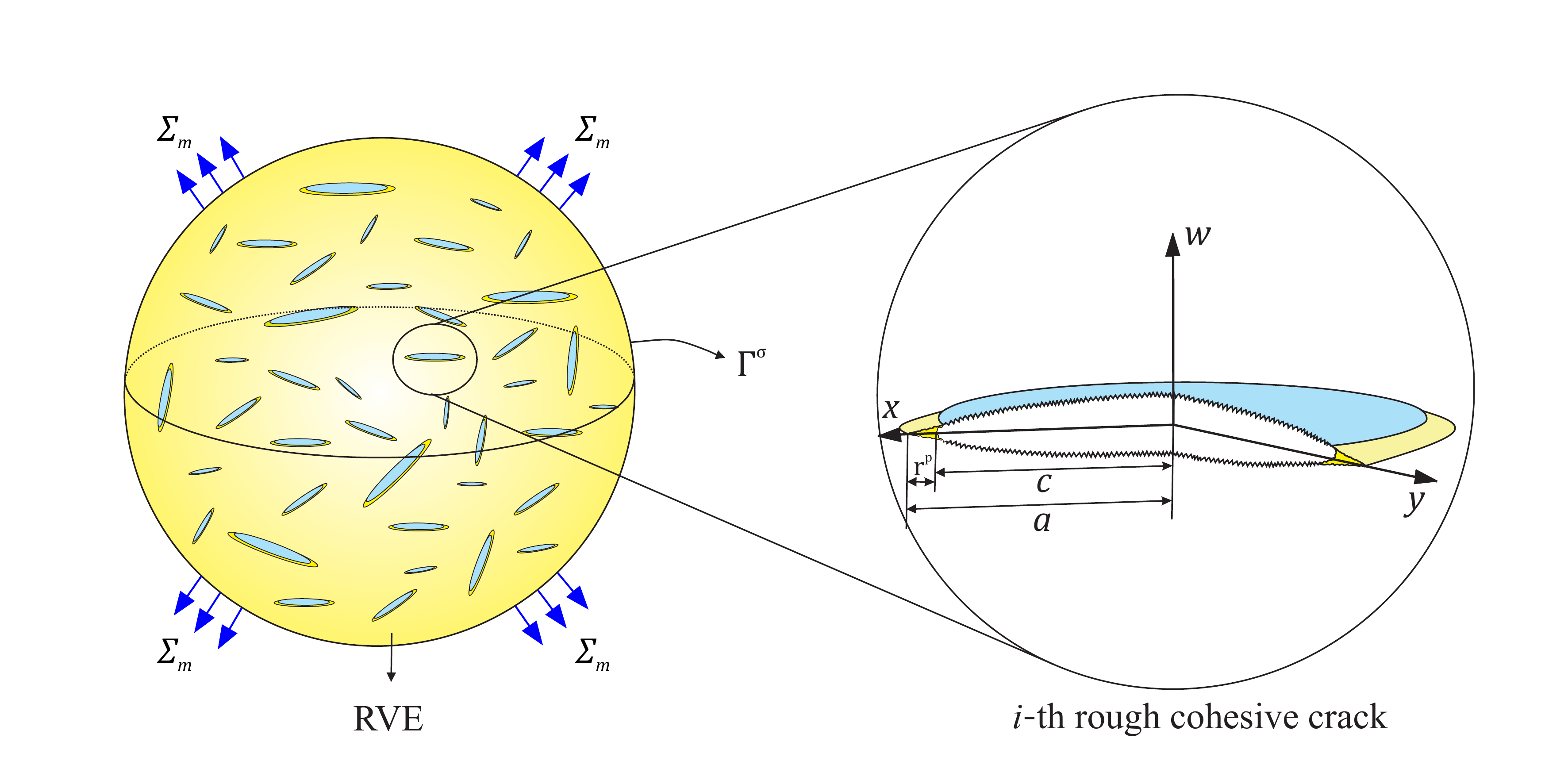}
		\caption{An isotropic medium containing randomly distributed microcracks of various sizes and orientations.}
		\label{RoPeShCr}
	\end{center}
\end{figure*}

The closed-form solution of a fractal penny-shaped cohesive crack (Fig. \ref{RoPeShCr}) is not a mathematically tractable problem, so what is heuristically tried here is implementing an analogy between 2-D and 3-D cohesive cracks to derive analytical expressions of interest. The basis of this analogy is special feature of fracture surfaces. As the crack advances, the corresponding fracture surface can be categorized into three distinct regions based on the degree of roughness. The consequent regions are named  ``Mirror’’, ``Mist’’, and ``Hackle’’, and this observation of roughness increase is named Mirror-Mist-Hackle (MMH) phenomenon. Several experimental research have reported observations of MMH phenomenon (see \cite{quinn1999,doquet2013,chopra2018}). It is observed that, in the case of penny-shaped cracks, the surface roughness increases in a radial direction away from the fracture initiation point, so the mirror striae are parallel to the direction of crack propagation and pass through the Hackle lines \cite{bradt2011}. A theoretical proof of the MMH phenomenon is given in the work of Wnuk and Yavari \cite{wnuk2008}. They suggested a relationship between the fractal exponent and the normalized crack length.

\begin{figure}
	\begin{center}
		\includegraphics[width=0.5\linewidth]{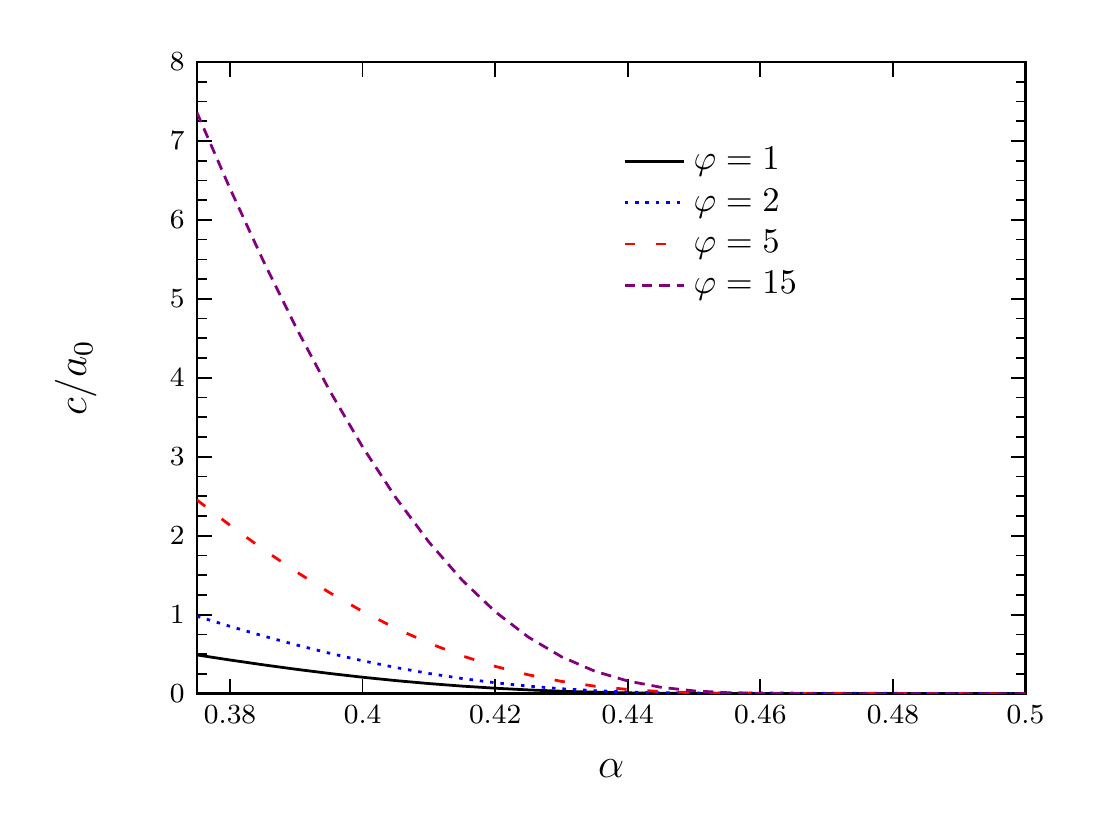}
		\caption{Dependence of the fractal exponent on the current normalized crack length.The parameter that distinguishes the curves shown is the microstructural constant $\varphi = \rho_{\alpha}/a_0$, where $a_0$ denotes the fracture quantum. It is seen that for short cracks, when $\varphi \longrightarrow0$ the function describing the Mirror\textendash Mist\textendash Hackle effect
			becomes extremely sensitive to the variations in the fractal exponent $\alpha$.}
		\label{crack length and roughness}
	\end{center}
\end{figure}

After some algebra and implementation of the proposed methodology in \cite{khezrzadeh2011}, the following modified relation between the fractal exponent and nominal crack length can be established,
\begin{equation}\label{CLRD}
\dfrac{\rho_\alpha}{c}=\dfrac{1}{\pi}\Big[\dfrac{\xi(\alpha)}{2^{1+\alpha}(0.05)^{\alpha}}   \Big]^{\dfrac{2}{2\alpha -1}},
\end{equation}
where $ \rho_\alpha $ is the radius of curvature of the equivalent smooth blunt crack, and $\xi(\alpha)=\dfrac{\pi^{\alpha-1}\Gamma(\alpha)}{\Gamma(\alpha+1/2)} $ (see Fig. \ref{crack length and roughness}).
Experimental measurements also indicate the existence of the MMH phenomenon. Arakawa and Takahashi \cite{arakawa1991} measured the roughness of the crack surface in the direction perpendicular to the direction of crack propagation at intervals of $ 0.5 mm $. Their measurements showed very close surface roughness at each interval, whereas this roughness increases with crack advancement. Therefore, building an analogy between the slit-like and penny-shaped cracks would be appropriate. The analogy is built to prevent the mathematical complexity raised when dealing with non-Euclidean objects. A fractal penny-shaped cohesive crack is produced by revolving a 2-D  fractal cohesive crack. In this analogy, the cross-section of the fractal penny-shaped crack is a slit-like crack. In other words, the displacement of a cross-section of this crack can be approximated with a Barenblatt-Dugdale slit-like crack. It is needed to mention that the analogy represented in this research entirely agrees with the case of smooth crack, and it has been just extended to the case of rough crack. In addition, the roughness advancement of a three-dimensional crack surface in the radial direction is an appropriate justification of the analogy between a rough 2-D crack and a rough 3-D crack.

Special care must be taken into account when the analogy is constructed. In the 2-D crack case, the asymptotic behavior of the surface energy of fracture surface is $ l^{1/H} $ (i.e., $ r=\sqrt{x^2+y^2}\sim l $). When the analogy is used, the fracture surface energy of 3-D crack behaves as $l^{1/H+1}$. On the other hand, without applying the conversion, the surface energy for 3-D crack propagation asymptotically behaves as $ \bigtriangleup U_s=2A\gamma = l^{2/H} $. So, if $H_1$ and $H_2$ denote the Hurst exponents of the conversion and true cases, receptively, then $H_2=\dfrac{2H_1}{H_1+1}$, where $ \dfrac{2}{3}<H_2<1 $ (see \cite{yavari2002} for more details).

To build this analogy, the size of the cohesive zone of slit-like crack must be equal to the size of the cohesive zone of a penny-shaped cohesive crack. Furthermore, the corresponding energy release rate for a cross-section of penny-shaped cohesive crack must be equal to the energy release rate of  Barenblatt\textendash Dugdale slit-like crack. Implementing an asymptotic analysis of the displacements of the penny-shaped and slit-like cohesive cracks in the close vicinity of the physical crack tips lead to the following crack edge opening displacements for a cross-section of the penny-shaped cohesive crack,
\begin{eqnarray}
   \delta_1&=&2u_w(r\longrightarrow c)\nonumber\\
   &=& \frac{4}{\pi}(\frac{1-\nu}{\mu})\lim_{r\longrightarrow c}\Bigg(\sigma^{\infty} \sqrt{a^2-r^2}\nonumber
   -\sigma_0 \int_r^a \frac{\sqrt{2c}\sqrt{t-c}+\dfrac{(t-c)^{3/2}}{2\sqrt{2c}}+O[t-c]^{5/2}}{\sqrt{2r}\sqrt{t-r}+\dfrac{(t-r)^{3/2}}{2\sqrt{2r}}+O[t-r]^{5/2}} dt\Bigg)\nonumber \\
   &=&\dfrac{4}{\pi}\left( \dfrac{1-\nu}{\mu}\right) \sigma_0 \dfrac{c (a-c)}{a},
\end{eqnarray}
and for the slit-like cohesive crack,
\begin{eqnarray}\label{slit-like J-integral}
    \delta_2&=&\dfrac{4}{\pi} \dfrac{1}{\mu (1+\nu)} \sigma_0 c \mathrm{ln}\left( \dfrac{a}{c} \right)=\dfrac{4}{\pi} \dfrac{1}{\mu (1+\nu)} \sigma_0 c \Big( (1-\dfrac{c}{a})+O[1-\dfrac{c}{a}]^2 \Big)\nonumber\\
    &=&\dfrac{4}{\pi} \dfrac{1}{\mu (1+\nu)} \sigma_0\dfrac{c (a-c)}{a},
\end{eqnarray}
respectively. The following ratio between $ J $-Integrals of these cracks is defined:
\begin{equation}
    \gamma=\dfrac{J_2}{J_1}=\dfrac{\sigma_0 \delta_2}{\sigma_0 \delta_1}=1-\nu^2.
\end{equation}
By applying this ratio in the displacement of slit-like crack, displacement of a cross-section of penny-shaped cohesive crack can be approximated. Note that the relation between the applied stresses of the 2-D and 3-D problems is
\begin{equation}
	\sigma_{\infty}^{(2D)}=\dfrac{2 \sigma_0}{\pi} \sin^{-1}\left( \dfrac{\sigma_{\infty}^{(3D)}}{\sigma_0}\right).
\end{equation}
So, by using Eq.~(\ref{cohesive size of penny}), the displacement field of a smooth penny-shaped crack is obtained as
\begin{equation}
     v^{s}_p(r)=(\frac{1-\nu}{\mu}) \frac{2\sigma_0}{\pi} \bigg[c \coth^{-1}\left( \sqrt{\frac{a^2-r^2}{a^2-c^2}} \right)-r \coth^{-1}\left( \frac{c}{r} \sqrt{\frac{a^2-r^2}{a^2-c^2}}\right) \bigg]   , \hspace{2cm} 0\leqslant r\leqslant a.
\end{equation}
The same pathway is used to deduce the analytical relations for a fractal penny-shaped crack from a fractal slit-crack. 

\begin{figure}[t]
	\begin{center}
		\includegraphics[width=0.5\linewidth]{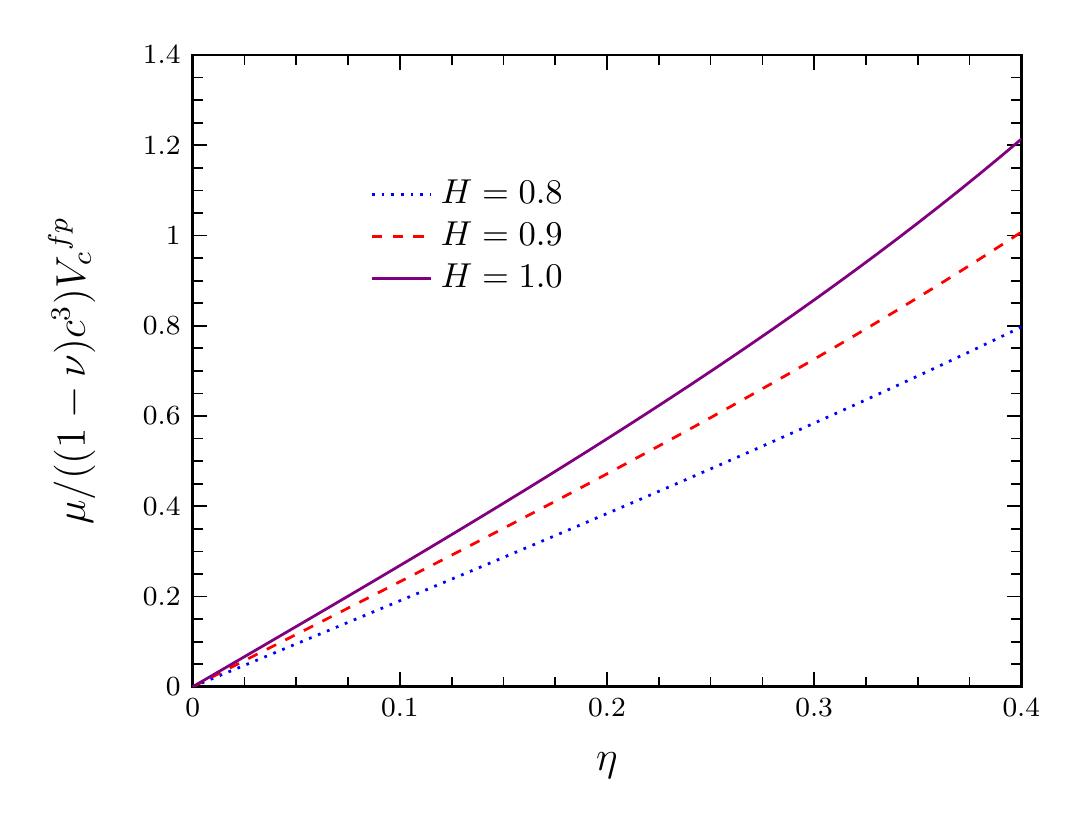}
		\caption{The volume of crack opening of the fractal penny-shaped crack for different roughness exponents ($ c=1 $).}
		\label{CVP}
	\end{center}
\end{figure}

Crack opening displacement (COD) in cohesive fractal crack can be approximated by superposing a fractal crack of finite nominal length $ 2a $  and a semi-infinite fractal crack. It is noteworthy that this approximation is acceptable in small-scale yielding, i.e., when $ r^p \ll c $, where $ r^p $ is the size of the cohesive zone. Moreover, a small cohesive zone versus crack length satisfies Barenblatt's condition, where there are no restrictions imposed on material ductility. The Westergaard function of a fractal crack of finite length $ 2a $  can be expressed as \cite{khezrzadeh2011}
\begin{equation}
Z^{(c)}(z;\alpha)=\frac{\sigma^{\infty}}{e^{i(1/2-\alpha)2|\pi /2-\theta |}}\frac{z^{2\alpha}}{(z^2-a^2)^{\alpha}}.
\end{equation}
Solution for the cohesive part of fractal crack can be obtained from the result for a semi-infinite fractal crack which is loaded by a pair of uniform loads of magnitude $ \sigma_0 $ along the segment $ c<x<a $ at the crack tip \cite{khezrzadeh2011},
\begin{eqnarray}
Z^{(s)}(z;\alpha)=-\frac{\sigma_0}{e^{i(1/2-\alpha)\theta}}\frac{1}{\pi (1+\alpha)}\bigg(\frac{a-c}{z-a}\bigg)^{1+\alpha}
 f_1\left(1,1+\alpha ,2+\alpha , -\frac{a-c}{z-a}\right),
\end{eqnarray}
where $ f_1 $ is hypergeometric function \cite{weisstein2002}.
Consequently, the Westergaard stress function of a fractal Barenblatt\textendash Dugdale crack can be established as,
\begin{equation}\label{Z of fractal Barenblatt-Dugdale}
Z^{t}(z;\alpha)=Z^{(c)}+Z^{(s)}.
\end{equation}
The following relation for the size of the cohesive zone of a rough crack can be read to vanish the stress intensity factor $K_I^f$ at the crack tip:
\begin{equation}
    r^p=cM=c (2\alpha)^{\frac{1}{\alpha}} \left( \dfrac{1}{\sqrt{1-\eta^2}}-1\right) (\sin^{-1} \eta)^{1/\alpha-2},
\end{equation}
where $\eta=\dfrac{\sigma^{\infty}}{\sigma_0}$. The yield zone stress $\sigma_0$ may not be uniform in the cohesive region. In this study, due to small-scale yielding, without loss of generality, it is considered to be uniform.
Integrating Eq.~(\ref{Z of fractal Barenblatt-Dugdale}) leads to:
\begin{eqnarray}
      \bar{Z}^{t}&=& -\frac{\sigma_0 \sin^{-1}\left( \dfrac{\sigma^{\infty}}{\sigma_0}\right) a}{e^{i(1/2-\alpha)2|\pi /2-\theta |}  \pi i^{2\alpha}}
     f_2\left(1-\dfrac{z^2}{a^2},1-\alpha,\dfrac{1}{2}+\alpha \right) \nonumber \\
   		&-&\frac{\sigma_0 (a-c) \Gamma (\alpha)}{\pi (1+\alpha) \Gamma (1+\alpha)} \left( \frac{a-c}{z-a}  \right)^{\alpha} f_1 \left(1 , \alpha , 2+\alpha , -\frac{a-c}{z-a}\right).
\end{eqnarray}
where $f_2$ gives the incomplete beta function \cite{weisstein2002}.
Therefore, the vertical opening displacement of the fractal penny-shaped crack can be written in the following form:
\begin{equation}
      v^{f}_p=\frac{1-\nu}{\mu} \mathrm{Im}(\bar{Z}^{t}).
\end{equation}
The COD and the cohesive zone size decrease as the roughness increases.
The volume of crack opening $ V_c^{fp} $ of a fractal microcrack can be calculated as
\begin{eqnarray}\label{Vc of fractal p-sh crack}
      V_c^{fp}&=&\int_{\Omega} [v^{f}_p] dV = \int_0^{2\pi} \int_0^a 2 v^{f}_p rdrd\theta=\frac{1-\nu}{\mu}2\pi c^3 \sigma_0 g_1(\eta ; \alpha).
\end{eqnarray}
where $\Omega$ denotes the surface of the extended penny-shaped crack.
In this equation, $g_1(\eta ; \alpha)$ is a complicated function of roughness exponent and stress ratio, which can be approximated by a function with monomial and exponential terms to capture regular or high gradient behavior like
\begin{eqnarray}
g_1(\eta ; \alpha) =b_1+b_2 \eta +\dfrac{b_3}{1- \eta} + \dfrac{b_4}{(1-\eta)^2},
\end{eqnarray}

where constants $b_i$ are obtained for different roughness exponents in Table \ref{constants1}, and can be used to calculate the volume of crack opening. The volume of the crack opening is plotted in Fig. \ref{CVP} for different values of roughness exponents. The values of $\eta$ are restricted to those which guarantee small-scale yielding. Considering $\eta <0.4 $, will result in having a ratio of the yielding zone's size to crack radius of less than $0.1$. As expected Eq.~(\ref{Vc of fractal p-sh crack}) predicts smaller $ V_c^{fp} $ for a lower roughness exponent $H$.
\begin{table}[h!]
	\begin{center}
		\caption{The constant coefficients of $g_1$ for some roughness exponents. $R^2$ represents the coefficient of determination.}
		\label{constants1}
		\begin{tabular}{l|c|c|c|c|c}
			$H$ & $b_1$ & $b_2$ & $b_3$ & $b_4$ & $R^2$ \\
			\hline
			0.80 & -0.036 & 0.267 & 0.036 & -0.0009 & $0.99$ \\
			0.85 & -0.054 & 0.275 & 0.054 & -0.0006 & $0.99$ \\
			0.90 & -0.076 & 0.272 & 0.076 & 0.0002 & $0.99$ \\
			0.95 & -0.101 & 0.257 & 0.102 & 0.0016 & $0.99$ \\
			1.00 & -0.124 & 0.237 & 0.125 & 0.0030 & $0.99$ \\
		\end{tabular}
	\end{center}
\end{table}

\section{Effective elastic material properties of an RVE containing rough microcracks}
\label{sec3}
The amount of dissipated energy during crack propagation is an essential step in the damage process modeling. This is not a straightforward task due to cohesive fracture and plastic dissipation in the cohesive zone. An upper bound for energy consumption in the damage process can be estimated by considering that total energy is consumed in the surface separation process. The total energy release due to a fractal penny-shaped crack in an RVE that is subjected to a triaxial uniform tension field can be estimated as,
\begin{equation}\label{Gf}
        G^f=\int_{\Omega} \sigma^{\infty} [v^{f}_p] dS -\int _{\Omega_y} \sigma_0 [v^{f}_p] dS= ( \sigma^{\infty} V_c^{fp} - \sigma_0 V_c^{'} ) ,
\end{equation}
where the values for $V_c^{'}$ can be reached from following relation: 
\begin{equation}
     V_c^{'} =\frac{1-\nu}{\mu}2\pi c^3 \sigma_0 g_2(\eta ; \alpha),
\end{equation}
with having $g_2(\eta ; \alpha)$ defined from the following correlation function, 
\begin{equation}
	g_2(\eta ; \alpha) =b_5+b_6 \eta + \dfrac{b_7}{1-\eta}+\dfrac{b_8}{(1-\eta)^2}.
\end{equation}
The coefficients of the above function are presented in Table \ref{constants2}. In Eq.~(\ref{Gf}), $\Omega_y$ denotes cohesive zone-yielded ring. Note that, as an approximation, the other processes which participate in the process of energy dissipation are not included.

\begin{table}[h!]
	\begin{center}
		\caption{The constant coefficients of $g_2$ for some roughness exponents.  $R^2$ represents the coefficient of determination.}
		\label{constants2}
		\begin{tabular}{l|c|c|c|c|c}
			$H$ & $b_5$ & $b_6$ & $b_7$ & $b_8$  & $R^2$ \\
			\hline
			0.80 & -0.0067 & -0.018 & 0.008 & -0.0002 & 0.99 \\
			0.85 & -0.0118 & -0.034 & 0.014 & 0.0002 & 0.99 \\
			0.90 & -0.0191 & -0.059 & 0.022 & 0.0013 & 0.99 \\
			0.95 & -0.0293 & -0.094 & 0.033 & 0.0029 & 0.99 \\
			1.00 & -0.0397 & -0.129 & 0.045 & 0.0045 & 0.99 \\
		\end{tabular}
	\end{center}
\end{table}

It is proven that in ductile materials, the damage is associated with void growth, while in the case of brittle and quasi-brittle bodies, the fracture is initiated from the cracks. Consequently, in the latter, critical cracks significantly influence the strength of the body. The total crack volume fraction of RVE is defined as the density of spheres encircling the cracks of the RVE, i.e.,
\begin{equation} \label{f}
f_{\Omega}:=\sum_{n=1}^{N_c}\frac{4\pi}{3V} c_n^3 \beta,
\end{equation}
where $ c_n $ and $ N_c $, are the radius of $ n $-th crack and the number of penny-shaped cracks in RVE, respectively, and $ \beta $ is ratio of volume of the permanent crack opening to volume of the total crack opening of a cohesive crack (i.e., $ 0\leqslant \beta \leqslant 1 $).
The crack opening volume fraction $ f_{\Omega} $ depends on the crack aspect ratio $ X\equiv \dfrac{[u_z]_{max}}{c} $, i.e., $ X\ll 1 $, which is a function of applied loading. It is concluded that while the cracks are open, load-induced changes of $ X $ do not have a significant effect on the micromechanical stiffness estimates \cite{dormieux2006}.
The portion of each crack length in $f_{\Omega}$ can be obtained using appropriate crack size distribution. Here, two common distribution functions, the Weibull and Log-Normal distributions, are chosen and defined as \footnote{ $\mu_g$ and $\sigma_g^2$ are mean and standard deviation of the variables natural logarithm. If mean and variance of normal crack distribution are $\mu_x$ and $\sigma_x^2$, one can use $\mu_g=\log \mu_x^2/\sqrt{\mu_x^2+\sigma_x^2}$ and $\sigma_g^2=\log \left( 1+\sigma_x^2/\mu_x^2 \right)$. So, $\mu_g$ and $\sigma_g>0$ are two real numbers.}
\begin{equation}
\label{LNDist}
\mathcal{LN}(x|\mu_g,\sigma_g)=\dfrac{1}{x \sigma_g \sqrt{2\pi}}e^{-\dfrac{(\log x-\mu_g)^2}{2\sigma_g^2}},
\end{equation}
and
\footnote{ $\kappa_g>0$ and $\lambda_g>0$ are the shape parameter and the scale parameter of the Weibull distribution, respectively.}
\begin{equation}
\label{WDist}
\mathcal{W}(x|\kappa_g,\lambda_g)=\dfrac{\kappa_g}{\lambda_g}\left( \dfrac{x}{\lambda_g}\right)^{\kappa_g - 1} e^{-\left(\dfrac{x}{\lambda_g}\right)^{\kappa_g}}.
\end{equation}
The well-known Weibull distribution was implemented to represent the size distribution of cracks in several research (\cite{trustrum1979, matsuo1986, matsuo1987, danzer2007}). 
Since the crack roughness is a function of its length, each length scale differently contributes to the overall material property. The crack size distribution function is defined on a continuous domain that can be discretized to a definite number of intervals. The volume fraction of cracks at each of these intervals is denoted by $ f_\Omega^{(i)} $. Each step covers an interval of the complete crack size distribution, and the average length in the step is the representative length of that interval, namely $ c^{(i)} $. The representative roughness of each length scale interval can be obtained from Eq.~(\ref{CLRD}). 
Consequently, the material properties can be reached by implementing a multi-step method \cite{khezrzadeh2016,khezrzadeh2017}, in which effective properties at each step corresponding to an interval are calculated by considering RVE with only $ f_\Omega^{(i)} $. The results at each step of self-consistent homogenization are used as the matrix properties of the next homogenization step. In this scheme, the process starts from the lowest length scale, and the effective material properties of the step are implemented as the next step's matrix properties. This way is continued until the whole length scales are involved in the analysis. The process is depicted in Fig.~\ref{MS_RVE} 

\begin{figure}[h]
	\begin{center}
		\includegraphics[width=0.5\linewidth]{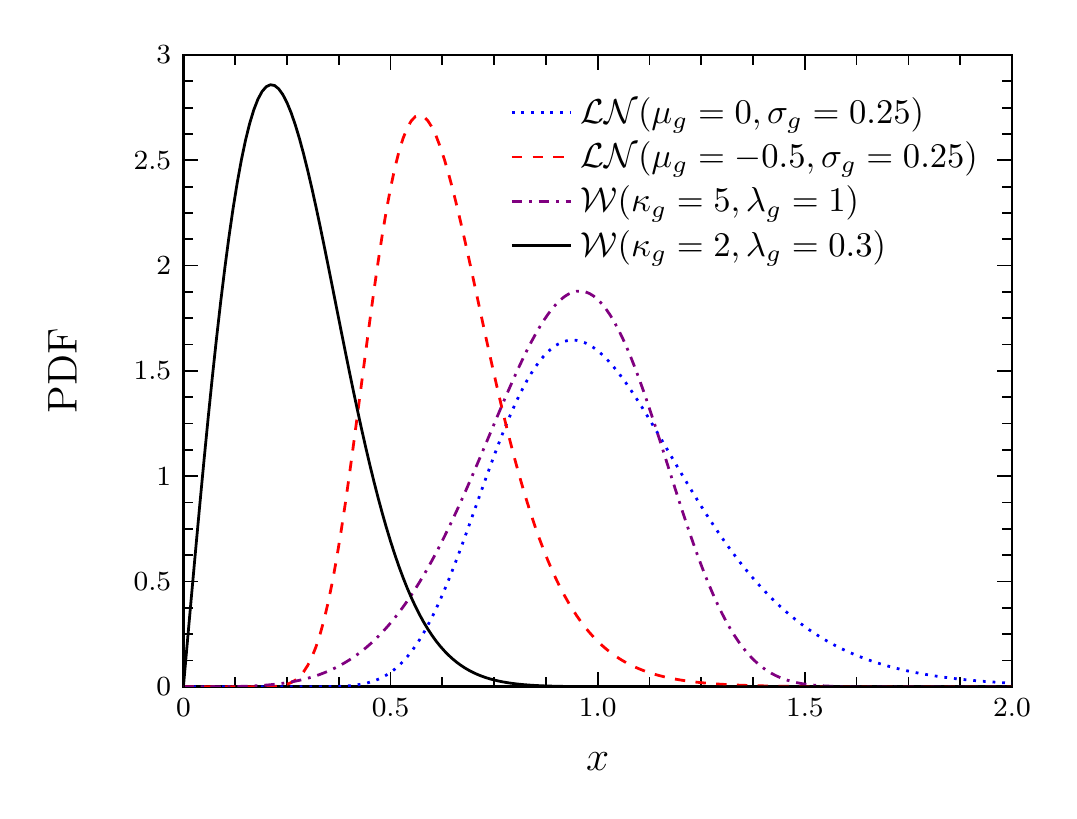}
		\caption{some PDFs used to represent cracks distribution in media }
		\label{DFunc}
	\end{center}
\end{figure}

In order to give a dimensionless representation, the crack sizes are normalized by defining parameter $x$ as: $x=\dfrac{a}{a_c}$ ($a_c$ corresponds to the smallest crack size that its roughness reaches to the threshold $H=0.8$), and the range of the normalized crack size is considered to be $x\in [x_{min},x_{max}]$. This normalization can exclude the dependency of $x$ on $\varphi$ as presented in Eq.~(\ref{CLRD}).
Then, considering $m$ steps, the normalized representative length and the size range of the $i$-th step are, $x^{(i)}$ and $s=(x_{max}-x_{min})/m$, respectively. Note that the roughness corresponded to $x^{(i)}$ is allocated for all crack sizes within $[x^{(i)}-s/2,x^{(i)}+s/2]$. Now, by defining the crack size distribution by a proper probability distribution function $P(x|\beta_1,\beta_2)$ (where $\beta_1$ and $\beta_2$ are distribution parameters), the crack volume fraction at $i$-th length scale can be calculated as

\begin{equation}
f_{\Omega}^{(i)}=f_{\Omega}\dfrac{\int_{x^{(i)}-s/2}^{x^{(i)}+s/2} x^3 P(x|\beta_1,\beta_2) d x}{\int_{0}^{\infty} x^3 P(x|\beta_1,\beta_2) d x-f_{\Omega} \int_{0}^{x^{(i)}-s/2} x^3 P(x|\beta_1,\beta_2) d x}
\end{equation}
Hereafter, a statistical representation is needed to give information about the size distribution of cracks. The well-known distribution of Weibull and Log-Normal distribution are chosen in this study. Sample probability distribution function (PDF) plots of these functions are given in Fig.~\ref{DFunc}. It should be noted that, for cracks with the normalized crack sizes larger than 1 ($x>1$), the roughness reached the ultimate value of $H=0.8$ as predicted in \cite{yavari2010}, and the roughness for other length scales can be calculated from Eq.~(\ref{CLRD}). Thenceforth, $\rho_{\alpha}$ is considered same for all length scales (see \cite{wnuk2008} for more details).

To construct the statistical multi-step self-consistent (SMSSC) micromechanical model, it is needed to have the energy release density in the $i-$th length scale. It can be calculated as:
\begin{eqnarray}
\frac{G^{f(i)}_t}{V}=\frac{3(1-\nu) f_{\Omega}^{(i)}}{2\beta \overline{\mu}^{(i)}} \sigma_0 \bigg[\sigma^{\infty} g_1(\eta ; \alpha) -\sigma_0 g_2(\eta ; \alpha) \bigg].
\end{eqnarray}
where $\overline{\mu}^{(i)}$ is the effective shear modulus of $i-$th length scale. 
The energy is dissipated in the process of cohesive fracture in complicated phenomena, and it can be assumed that this energy is consumed in the process of surface separation and plastic dissipation (see \cite{kfouri1979,mura1987,wnuk1990}). To obtain the elastic macro-strain tensor, it is needed to approximate the overall complementary energy density in the RVE. Hence, the sum of the uncracked body's complementary energy density ($W$), and the density of energy release due to microcracks distribution, is used to estimate the overall complementary energy density. This can be expressed as follows,

\begin{equation}
          W^{(i)} = W^{(i-1)}+ \frac{G^{f(i)}_t}{V}=\dfrac{1}{2}\sigma^{\infty}_{ij}D_{ijkl}^{(i-1)} \sigma^{\infty}_{kl}+\frac{3(1-\nu) f_{\Omega}^{(i)}}{2\beta \overline{\mu}^{(i)}}
           \sigma_0 \bigg[\sigma^{\infty} g_1(\eta ; \alpha) -\sigma_0 g_2(\eta ; \alpha) \bigg] .
\end{equation}

Additionally, the effective elastic compliance moduli in $i-$th step ($ \overline{\mathbb{D}}^{(i)} $) is the sum of elastic compliance of background material (the previous step) ($ \overline{\mathbb{D}}^{(i-1)} $) and the added compliance due to microcracks of the $i-$th step ($ \mathbb{H}^{(i)} $) \cite{li2004}. Thus,
\begin{equation}\label{EEC}
 \overline{\mathbb{D}}^{(i)}= \overline{\mathbb{D}}^{(i-1)}+\mathbb{H}^{(i)}.
\end{equation}
The macro strain $\varepsilon_{ij}^{(i)}$ can be derived from the overall complementary energy density $W^{(i)}$ of an RVE such as, 
\begin{eqnarray} \label{fractal macro strain}
\varepsilon_{ij}^{(i)} &=&\frac{\partial W^{(i)}}{\partial {\sigma^{\infty}_{ij}}}= \frac{\partial W^{(i-1)}}{\partial {\sigma^{\infty}_{ij}}}+ \frac{\partial (G^{f(i)}_t/V)}{{\sigma^{\infty}_{ij}}}\nonumber\\
&=&D_{ijkl}^{(i-1)} \sigma^{\infty}_{kl} + \frac{\partial (G^{f(i)}_t/V)}{\partial \sigma^{\infty}}\frac{{\partial \sigma^{\infty}}}{\partial \sigma^{\infty}_{ij}},
\end{eqnarray}
where $\varepsilon_{ij}^{(i)} = \varepsilon^{(i-1)}_{ij} +\varepsilon^{add(i)}_{ij}$ and $\varepsilon^{(i-1)}_{ij} =D_{ijkl}^{(i-1)}\sigma^{\infty}_{kl}$.
Therefore, for a given crack opening volume fraction, the additional elastic macro-strain tensor is obtained as
\begin{equation}\label{ASFT}
      \varepsilon^{add(i)}_{f}=\mathbb{H}^{(i)}:\mathbf{\sigma}=\frac{\partial (G^{f(i)}_t/V)}{3 \partial \sigma^{\infty}}I^{(2)}.
\end{equation}
Using Eqs.~(\ref{EEC})\textendash(\ref{ASFT}) leads to the following expression,
\begin{equation}\label{BSR}
      \frac{1}{3\overline{K}^{(i)}}= \frac{1}{3{\overline{K}^{(i-1)}}}+\dfrac{1}{\sigma^{\infty}}\frac{\partial (G^{f(i)}_t/V)}{3\partial \sigma^{\infty}}.
\end{equation}
To guarantee the positive definiteness of the effective strain energy, the following relation should hold \cite{li2004, blal2012},
\begin{equation}\label{BSRC}
\frac{\overline{K}^{(i)}}{\overline{K}^{(i-1)}}=\frac{\overline{\mu}^{(i)}}{\overline{\mu}^{(i-1)}}.
\end{equation}
The corresponding result of Eq.~(\ref{BSRC}) is that $\overline{\nu}^{(i)}=\overline{\nu}^{(i-1)}$.
  
Consequently, the effective bulk and shear moduli can be derived by using  relation (\ref{BSR}) and imposing the condition of Eq.~(\ref{BSRC}) that is purposed as a closure condition for pure hydrostatic loading \cite{li2004}. The result is
	\begin{equation}\label{RESBF}
	  \dfrac{\overline{\mu}^{(i)}}{\overline{\mu}^{(i-1)}}=1-\dfrac{1-\nu^2}{\beta(1-2\nu)}f_{\Omega}^{(i)}\sigma_0 \left[  \dfrac{g_1}{\sigma^{\infty}}+\dfrac{\partial g_1}{\partial \sigma^{\infty}}-\dfrac{\sigma_0}{\sigma^{\infty}}\dfrac{\partial g_2}{\partial \sigma^{\infty}}\right].
	\end{equation}
In Eq.~(\ref{RESBF}), $i$  is the counter of steps which starts from 1, and $\overline{\mu}^{(0)}$ is equal to the shear modulus of uncracked body, i.e.,  $\overline{\mu}^{(0)}=\mu$. Also, the shear modulus of final step is denoted by $\overline{\mu}^{(m)}=\overline{\mu}$  where $m$ is the number of steps in the homogenization process of statistical multi-step self-consistent (SMSSC) micromechanical model. Finally, the effective shear modulus of the cracked media read as,
\begin{equation}\label{SMSSC}
	\overline{\mu}=\overline{\mu}^{(m-1)}\left[ 1-\dfrac{1-\nu^2}{\beta(1-2\nu)}f_{\Omega}^{(m)}\sigma_0 \left[  \dfrac{g_1}{\sigma^{\infty}}+\dfrac{\partial g_1}{\partial \sigma^{\infty}}-\dfrac{\sigma_0}{\sigma^{\infty}}\dfrac{\partial g_2}{\partial \sigma^{\infty}}\right]\right].
\end{equation}

\begin{figure}[h]
	\begin{center}
		\includegraphics[width=0.5\linewidth]{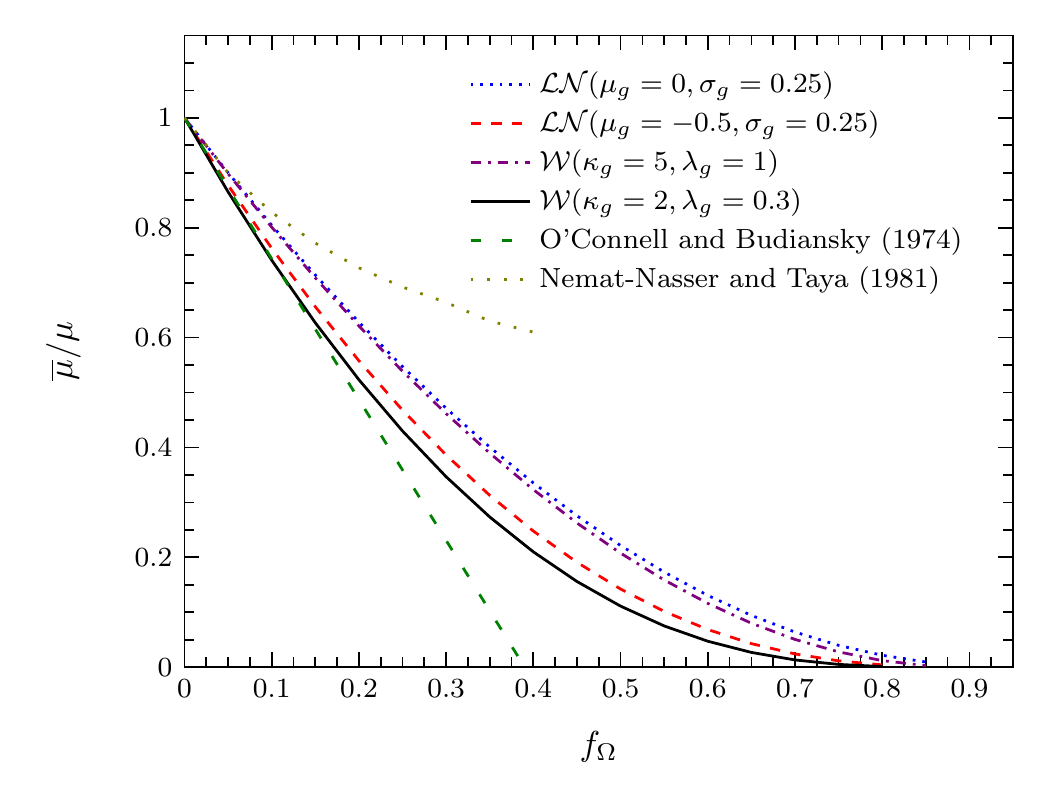}
		\caption{The effective shear modulus for some crack distributions ($ \dfrac{\sigma^{\infty}}{\sigma_0}=0.3 $, $ \nu=0.3 $). }
		\label{ESR}
	\end{center}
\end{figure}

As discussed, to calculate the homogenized properties of medium according to its crack size distribution, it is required to subdivide the range of crack size appropriately. The effective shear modulus in each step of iterations is obtained by considering the dilute distribution of microcracks to exclude the interaction effect.  Here, Weibull and Log-Normal distributions with distribution parameters presented in Fig.~\ref{DFunc} and $ f_{\Omega}^{(i)}\leqslant 0.05 $  are considered. The results of this homogenization scheme are plotted in Fig.~\ref{ESR} where it is evident that at the same crack volume, distributions with more portions of smaller cracks have lower effective shear modulus. This phenomenon directly is resulted from the fact that shorter cracks have higher CODs that contribute to comparatively increased compliance levels. In other words, among the four graphs shown in Fig.~\ref{DFunc}, $ \mathcal{W}(x|\kappa_g=2,\lambda_g=0.3) $ contains a big portion of smaller cracks, and consequently this distribution results in lower effective shear modulus.

When surfaces of the circular cracks are smooth, the energy release of a crack with reduce $a$ in an RVE of volume $V$ is equal to $\dfrac{8}{3} \dfrac{1-\nu}{\mu} \dfrac{a^3}{V} (\sigma^{\infty})^2$.  If the effective properties of the RVE is calculated in one step, the material properties reduce to those obtained by O'Connell and Budiansky \cite{OConnell1974} (see Fig. \ref{ESR}). A similar procedure was done by Kachanov and Sevostianov \cite{kachanov1992, kachanov2005} where they showed that for random orientation distribution of penny-shaped cracks, the energy release in the RVE can be calculated as:
\begin{equation}\label{KachanovEq}
\dfrac{G^f}{V}=\dfrac{4(1-\nu)}{9(1-\nu/2)\mu}\dfrac{\sum_i^N a^{(i)^2}}{V} [(1-\nu/2)\sigma_{ij}\sigma_{ji}-\dfrac{\nu}{10}(\sigma_{kk})^2]
\end{equation}
For the hydrostatic stress state, this expression reduces to the present study for $ H=1 $.
On the other hand, Nemat-Nasser and Taya \cite{nemat1981} calculated the effective elastic properties of a porous body containing periodically distributed voids by considering strain energy. The effective elastic moduli obtained by them are stiffer than the current study results due to the existence of less stress concentration at the boundary of spherical voids compared to the penny-shaped cracks. As seen in Fig.~\ref{ESR}, the effective material property of an RVE containing rough penny-shaped cracks falls between two bounds: the effective material properties of an RVE containing smooth penny-shaped cracks and an RVE containing spherical voids.

\section{The effective yielding potential function of an RVE containing rough cohesive cracks }
\label{sec4}
The homogenization of non-linear problems can be sometimes mathematically intractable, consequently some simplifying assumptions are needed. Hencky's maximum distortional energy principle is employed to establish a condition for yielding at the macroscopic level. According to this criterion, the distortional strain energy density has a limit, which is a material constant. So, the macroscopic yielding initiates when the distortional strain energy density reaches a threshold,
\begin{equation}
        U_d=\int_0^{\varepsilon_{ij}^{'}} \overline{\sigma}_{ij}^{'}d\overline{\varepsilon}_{ij}^{'} \leqslant U_d^{cr},
\end{equation}
where $ \overline{\sigma}_{ij}^{'} $ and $ \overline{\varepsilon}_{ij}^{'}=\dfrac{1}{2\overline{\mu}}\overline{\sigma}_{ij}^{'} $  are the deviatoric part of the macro-stress and the equivalent macro-deviatoric elastic strain tensors, respectively.
This criterion can be written as following,
\begin{equation}
         \dfrac{\sigma_{eq}^2}{\sigma_Y^2}=\dfrac{\overline{\mu}}{\mu}.
\end{equation}
Considering a multi-step self-consistent homogenization scheme and Eq.~(\ref{RESBF}), finally, the effective yielding potential function in term of the ratio $ \dfrac{\sigma_{eq}}{\sigma_Y} $ can be written in the following form:
\begin{equation}\label{Wang & Li yielding function}
     \Psi (\sigma_{eq}, \sigma^{\infty} , \textbf{q}) = \frac{\sigma_{eq}^2}{\sigma_Y^2} - \dfrac{\overline{\mu}}{\mu}=0,
\end{equation}
where $ \textbf{q} $ denotes the other internal variables which are implicitly embedded in $ \sigma_Y $ \cite{li2004}. $\dfrac{\overline{\mu}}{\mu}$ is the effective multi-step shear modulus resulted from Eq.~(\ref{SMSSC}).
It should be noted that, based on this relation, the constitutive relation in macro and micro levels are different. This model can predict material degradation and failure due to the growth of microcracks. The effect of Poisson's ratio on material damage is noticeable. Note that for incompressible materials (i.e., $ \nu=0.5 $) under uniform triaxial tension load, dilatational strain energy can not be reached, and the model will not give reasonable results. While the irreversible part of effective constitutive relation at the macro-level is characterized as pressure-sensitive plasticity, the reversible part of this relation has a nonlinear elasticity form.
\begin{figure}
	\begin{subfigure}{.5\textwidth}
		\centering
		\includegraphics[width=1\linewidth]{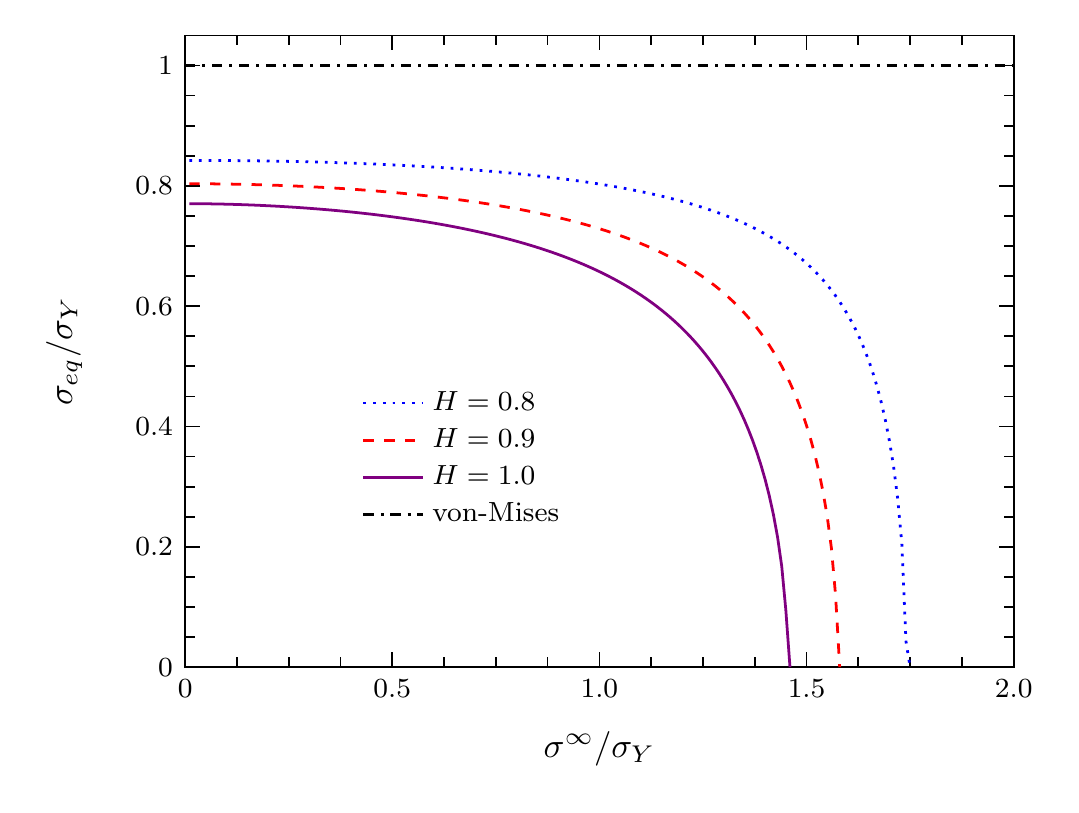}
		\caption{$ f_{\Omega}=0.1 $, $ \nu =0.2 $}
		\label{fig:sub-first}
	\end{subfigure}
	\begin{subfigure}{.5\textwidth}
		\centering
		\includegraphics[width=1\linewidth]{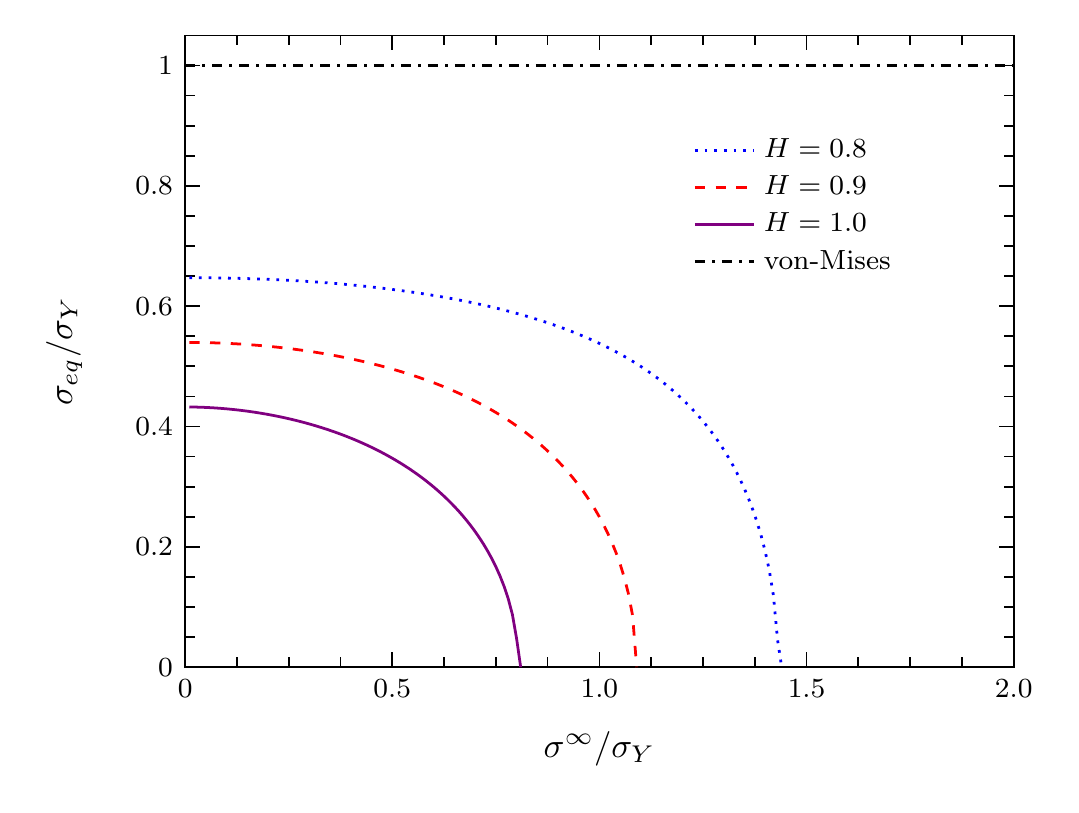}
		\caption{$ f_{\Omega}=0.2 $, $ \nu =0.2 $}
		\label{fig:sub-second}
	\end{subfigure}	
	\caption{The effective stress ratio for some crack surface roughnesses and von-Mises line}
	\label{fig:equivalent-stress-ratio}
\end{figure}
By considering the roughness limits, i.e., $ H=0.8 $  and $ H=1 $, one can obtain the upper and lower bounds of the yielding potential function (see Fig. \ref{fig:equivalent-stress-ratio}). Based on the results it can be stated that the yielding will occur at higher levels of stress for the bodies with rougher cracks. In other words, the strength of a material depends on the roughness of cracks within the body, it can be easily shown that this function reduces to Li and Wang \cite{li2004} yielding function for smooth penny-shaped cracks. One can postulate that material can endure higher amounts of $ f_\Omega $ when crack surface roughness reaches higher levels. It should be mentioned that this yield function will become $ J_2 $ (von-Mises line) for $ f_{\Omega}=0 $ and predicts a complete failure when $ \dfrac{\sigma_{eq}}{\sigma_Y} $ reaches a finite value even for infinitesimal amount of $ f_{\Omega} $. Note that COD is approximated with the sum of a fractal crack of finite length and a semi-infinite fractal crack. So, this model will be applicable in the cases of small-scale yielding.

\begin{figure}[h]
	\begin{subfigure}{.5\textwidth}
		\centering
		\includegraphics[width=1\linewidth]{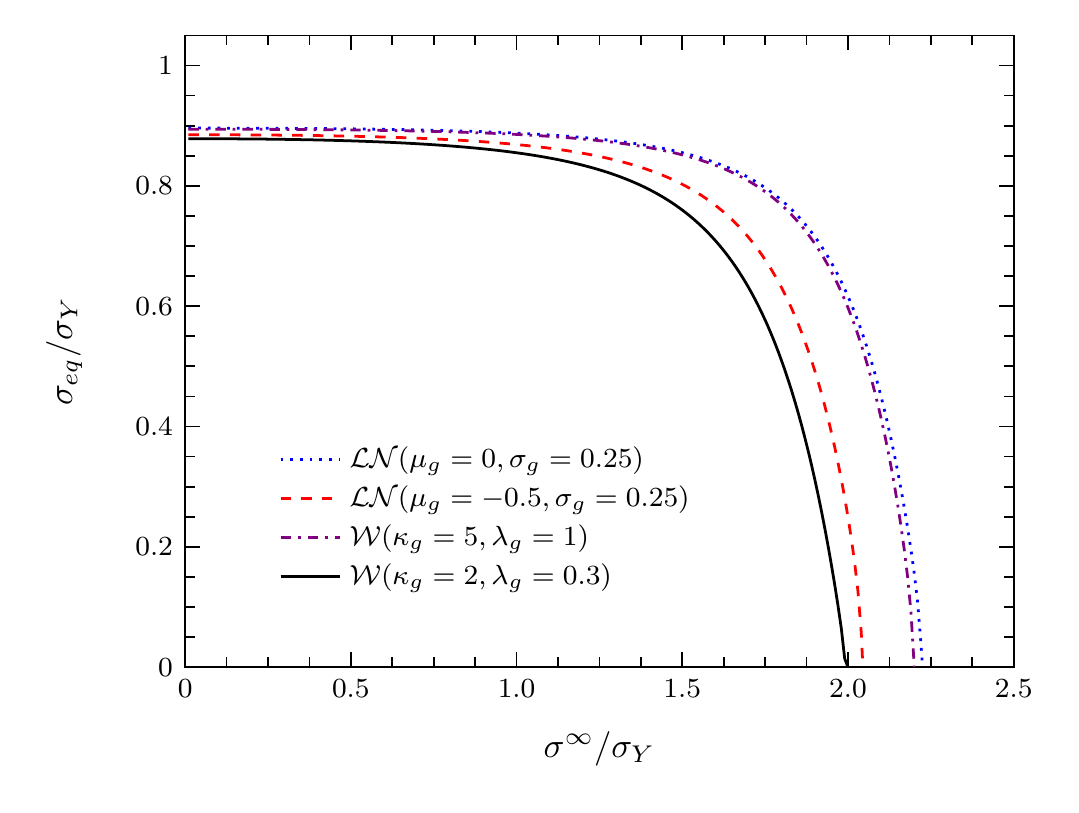}
		\caption{$ f_{\Omega}=0.1 $, $ \nu =0.3 $}
		\label{fig:sub-first-Multistep}
	\end{subfigure}
	\begin{subfigure}{.5\textwidth}
		\centering
		\includegraphics[width=1\linewidth]{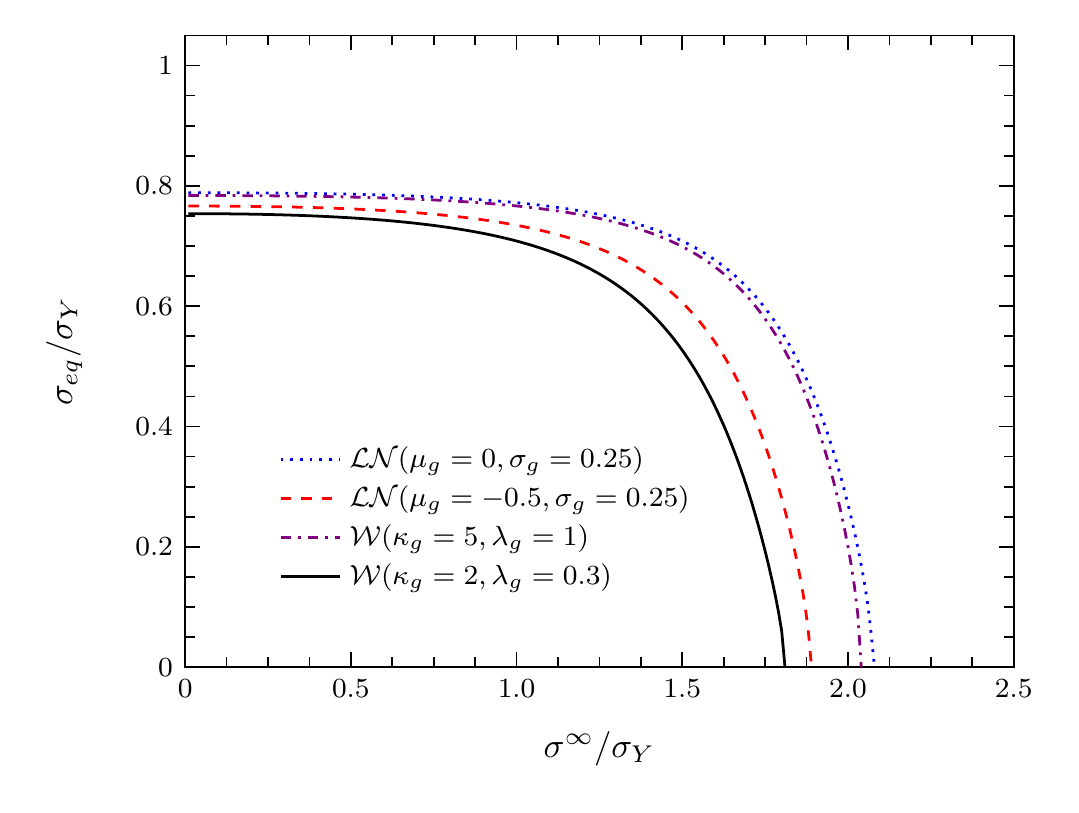}
		\caption{$ f_{\Omega}=0.2 $, $ \nu =0.3 $}
		\label{fig:sub-second-Multistep}
	\end{subfigure}	
	\newline	
	\begin{subfigure}{.5\textwidth}
		\centering
		\includegraphics[width=1\linewidth]{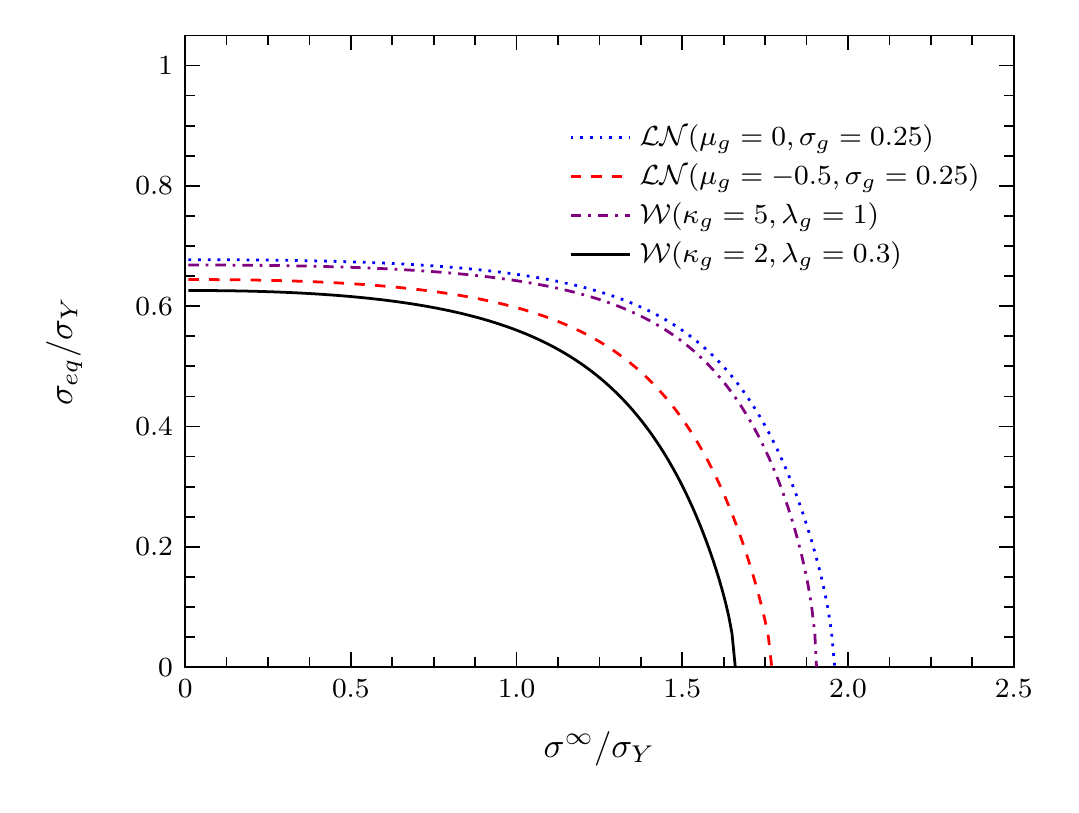}
		\caption{$ f_{\Omega}=0.3 $, $ \nu =0.3 $}
		\label{fig:sub-third-Multistep}
	\end{subfigure}
	\begin{subfigure}{.5\textwidth}
		\centering
		\includegraphics[width=1\linewidth]{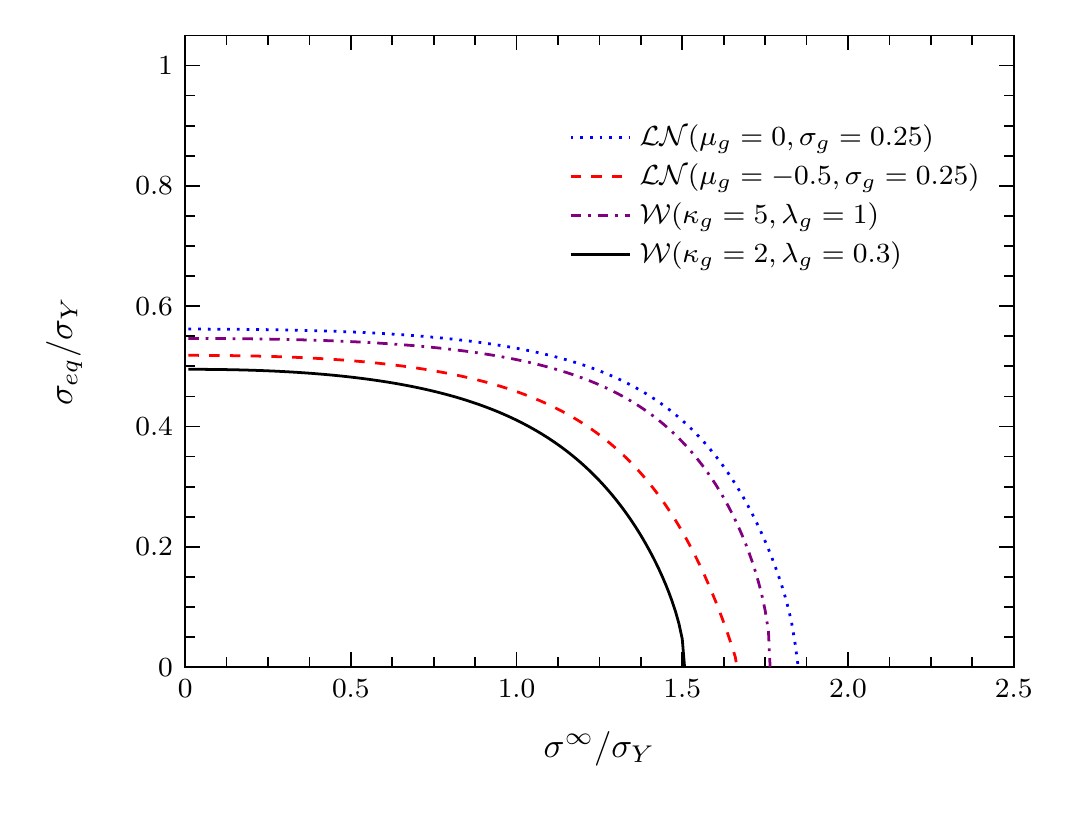}
		\caption{$ f_{\Omega}=0.4 $, $ \nu =0.3 $}
		\label{fig:sub-forth-Multistep}
	\end{subfigure}
	
	\caption{The effective stress ratio for some crack distributions}
	\label{FigESRM}
\end{figure}

As mentioned in the previous section, proper PDFs such as Weibull and Log-Normal (see Eqs.~(\ref{LNDist} and \ref{WDist}) can be implemented to represent cracks' size distribution in a solid. The continuity of crack sizes results in continuous roughness exponents between 0.8 and 1. The range of roughness is discretized in this approach, and then the multi-step method is implemented to obtain the effective yielding potential function of the medium. For instance, in Figs.~\ref{FigESRM}, the yielding potential function of solid of some specific crack volume fraction is depicted. The bodies that comparatively contain higher portion of rougher cracks, withstand elevated levels of applied stresses.


\section{Conclusions}
\label{sec5}
The statistical multi-step self-consistent (SMSSC) approach is introduced and used as a framework of analysis for solids with distributed rough micro-cracks in the present work. In this method, the roughness and size distribution are included in the calculation procedure to reach such solids' elastoplastic properties. At first, a fractal model for penny-shaped cohesive cracks is proposed. It is reached that the direct dependence of roughness and cracks opening displacement influences the cohesive zone's size. Since the roughness itself is a function of nominal length, the corresponding statistical distribution of nominal crack length affects the fractured media's overall behavior. These multi-lateral effects are investigated through the SMSSC framework in terms of elastic and plastic properties of the solids containing a distribution of rough cracks. The significant outcomes of the SMSSC can be concluded as

$\bullet$ The statistical distribution of crack length imposes a considerable impact on the elastic properties of fractured media. At the same volume fraction of cracks, solids containing a distribution with higher shorter cracks percentage exhibit higher material stiffness reduction.

$\bullet$ The reached homogenized effective elastic moduli of solids containing rough microcracks fall between two bounds. These bounds are defined through the effective material properties of two solids containing the same volume fraction of defects. The defects considered for the lower bound are smooth penny-shaped cracks, while the case of spherical voids is considered for the upper bound. As the roughness increases, these moduli tend to the effective elastic moduli of the solid weakened by spherical voids.

$\bullet$ The results for yielding potential function indicate that the statistical distribution of cracks directly influences the plastic behavior of cracked solids. At a certain volume fraction of cracks, bodies with a higher concentration of short cracks will undergo plastic yielding at lower applied stresses.

\section*{Acknowledgments}
\label{sec6}
The authors thank Professor Arash Yavari (School of Civil and Environmental Engineering, Georgia Institute of Technology) for valuable discussions and the technical guides.





\bibliographystyle{elsarticle-num}
\bibliography{references}

\begin{thebibliography}{10}
\expandafter\ifx\csname url\endcsname\relax
  \def\url#1{\texttt{#1}}\fi
\expandafter\ifx\csname urlprefix\endcsname\relax\def\urlprefix{URL }\fi
\expandafter\ifx\csname href\endcsname\relax
  \def\href#1#2{#2} \def\path#1{#1}\fi

\bibitem{fjar2008}
E.~Fjar, R.~M. Holt, A.~Raaen, P.~Horsrud, Petroleum Related Rock Mechanics,
  Elsevier, 2008.

\bibitem{gueguen2011}
Y.~Gu{\'e}guen, M.~Kachanov, Effective elastic properties of cracked rocks - an
  overview, in: Mechanics of crustal rocks, Springer, 2011, pp. 73--125.

\bibitem{bristow1960}
J.~Bristow, Microcracks, and the static and dynamic elastic constants of
  annealed and heavily cold-worked metals, British Journal of Applied Physics
  11~(2) (1960) 81.

\bibitem{walsh1965a}
J.~Walsh, The effect of cracks on the compressibility of rock, Journal of
  Geophysical Research 70~(2) (1965) 381--389.

\bibitem{walsh1965b}
J.~Walsh, The effect of cracks on the uniaxial elastic compression of rocks,
  Journal of Geophysical Research 70~(2) (1965) 399--411.

\bibitem{kovalenko1977}
Y.~F. Kovalenko, R.~Salganik, Fractured inhomogeneities and their influence on
  effective mechanical characteristics, Izv. Akad. Nauk SSSR. Mekh. Tverd.
  Tela~(5) (1977) 76--86.

\bibitem{nemat1981}
S.~Nemat-Nasser, M.~Taya, On effective moduli of an elastic body containing
  periodically distributed voids, Quarterly of Applied Mathematics 39~(1)
  (1981) 43--59.

\bibitem{roberts2002}
A.~Roberts, E.~J. Garboczi, Computation of the linear elastic properties of
  random porous materials with a wide variety of microstructure, Proceedings of
  the Royal Society of London. Series A: Mathematical, Physical and Engineering
  Sciences 458~(2021) (2002) 1033--1054.

\bibitem{hill1965}
R.~Hill, A self-consistent mechanics of composite materials, Journal of the
  Mechanics and Physics of Solids 13~(4) (1965) 213--222.

\bibitem{OConnell1974}
R.~J. O'Connell, B.~Budiansky, Seismic velocities in dry and saturated cracked
  solids, Journal of Geophysical Research 79~(35) (1974) 5412--5426.

\bibitem{budiansky1976}
B.~Budiansky, R.~J. O'Connell, Elastic moduli of a cracked solid, International
  Journal of Solids and Structures 12~(2) (1976) 81--97.

\bibitem{hoenig1979}
A.~Hoenig, Elastic moduli of a non-randomly cracked body, International Journal
  of Solids and Structures 15~(2) (1979) 137--154.

\bibitem{salganik1973}
R.~Salganik, Mechanics of bodies with many cracks, Mech. Solids 8~(4) (1973)
  135--143.

\bibitem{hashin1988}
Z.~Hashin, The differential scheme and its application to cracked materials,
  Journal of the Mechanics and Physics of Solids 36~(6) (1988) 719--734.

\bibitem{zimmerman1990}
R.~W. Zimmerman, Compressibility of Sandstones, Elsevier, 1990.

\bibitem{mori1973}
T.~Mori, K.~Tanaka, Average stress in matrix and average elastic energy of
  materials with misfitting inclusions, Acta Metallurgica 21~(5) (1973)
  571--574.

\bibitem{benveniste1986}
Y.~Benveniste, On the mori-tanaka's method in cracked bodies, Mechanics
  Research Communications 13~(4) (1986) 193--201.

\bibitem{horii1990}
H.~Horii, K.~Sahasakmontri, Mechanical properties of cracked solids: validity
  of the self-consistent method, in: Micromechanics and Inhomogeneity,
  Springer, 1990, pp. 137--159.

\bibitem{nemat1993b}
S.~Nemat-Nasser, N.~Yu, M.~Hori, Solids with periodically distributed cracks,
  International Journal of Solids and Structures 30~(15) (1993) 2071--2095.

\bibitem{shen2001}
L.~Shen, S.~Yi, An effective inclusion model for effective moduli of
  heterogeneous materials with ellipsoidal inhomogeneities, International
  Journal of Solids and Structures 38~(32) (2001) 5789--5805.

\bibitem{berryman2002}
J.~G. Berryman, S.~R. Pride, H.~F. Wang, A differential scheme for elastic
  properties of rocks with dry or saturated cracks, Geophysical Journal
  International 151~(2) (2002) 597--611.

\bibitem{kachanov1980}
M.~Kachanov, Continuum model of medium with cracks, Journal of the Engineering
  Mechanics Division 106~(5) (1980) 1039--1051.

\bibitem{kachanov1992}
M.~Kachanov, Effective elastic properties of cracked solids: critical review of
  some basic concepts, Applied Mechanics Reviews 45~(8) (1992) 304--335.

\bibitem{kachanov1993}
M.~Kachanov, Elastic solids with many cracks and related problems, Advances in
  Applied Mechanics 30 (1993) 259--445.

\bibitem{kachanov2005}
M.~Kachanov, I.~Sevostianov, On quantitative characterization of
  microstructures and effective properties, International Journal of Solids and
  Structures 42~(2) (2005) 309--336.

\bibitem{garbin1973}
H.~Garbin, L.~Knopoff, The compressional modulus of a material permeated by a
  random distribution of circular cracks, Quarterly of Applied Mathematics
  30~(4) (1973) 453--464.

\bibitem{garbin1975a}
H.~Garbin, L.~Knopoff, The shear modulus of a material permeated by a random
  distribution of free circular cracks, Quarterly of Applied Mathematics 33~(3)
  (1975) 296--300.

\bibitem{garbin1975b}
H.~Garbin, L.~Knopoff, Elastic moduli of a medium with liquid-filled cracks,
  Quarterly of Applied Mathematics 33~(3) (1975) 301--303.

\bibitem{hudson1980}
J.~Hudson, Overall properties of a cracked solid, Mathematical Proceedings of
  the Cambridge Philosophical Society 88~(2) (1980) 371--384.

\bibitem{hudson1986}
J.~Hudson, A higher order approximation to the wave propagation constants for a
  cracked solid, Geophysical Journal International 87~(1) (1986) 265--274.

\bibitem{hudson1990}
J.~Hudson, Overall elastic properties of isotropic materials with arbitrary
  distribution of circular cracks, Geophysical Journal International 102~(2)
  (1990) 465--469.

\bibitem{horii1983}
H.~Horii, S.~Nemat-Nasser, Overall moduli of solids with microcracks:
  load-induced anisotropy, Journal of the Mechanics and Physics of Solids
  31~(2) (1983) 155--171.

\bibitem{yu1995}
S.-W. Yu, X.-Q. Feng, A micromechanics-based damage model for
  microcrack-weakened brittle solids, Mechanics of Materials 20~(1) (1995)
  59--76.

\bibitem{mavko1978}
G.~M. Mavko, A.~Nur, The effect of nonelliptical cracks on the compressibility
  of rocks, Journal of Geophysical Research: Solid Earth 83~(B9) (1978)
  4459--4468.

\bibitem{grechka2006}
V.~Grechka, I.~Vasconcelos, M.~Kachanov, The influence of crack shape on the
  effective elasticity of fractured rocks, Geophysics 71~(5) (2006) D153--D160.

\bibitem{mandelbrot1983}
B.~B. Mandelbrot, The fractal geometry of nature, Vol. 173, WH freeman New
  York, 1983.

\bibitem{mandelbrot1984}
B.~B. Mandelbrot, D.~E. Passoja, A.~J. Paullay, Fractal character of fracture
  surfaces of metals 308 (1984) 721--2.

\bibitem{saouma1994}
V.~E. Saouma, C.~C. Barton, Fractals, fractures, and size effects in concrete,
  Journal of Engineering Mechanics 120~(4) (1994) 835--854.

\bibitem{saouma1990}
V.~E. Saouma, C.~C. Barton, N.~A. Gamaleldin, Fractal characterization of
  fracture surfaces in concrete, Engineering Fracture Mechanics 35~(1) (1990)
  47--53.

\bibitem{cherepanov1995}
G.~P. Cherepanov, A.~S. Balankin, V.~S. Ivanova, Fractal fracture mechanics—a
  review, Engineering Fracture Mechanics 51~(6) (1995) 997--1033.

\bibitem{mosolov1991}
A.~Mosolov, Cracks with fractal surfaces, Dokl Akad Nauk SSSR 319~(4) (1991)
  840--4.

\bibitem{balankin1996a}
A.~Balankin, A.~Bravo-Ortega, M.~Galicia-Cortes, O.~Susarey, The effect of
  self-affine roughness on crack mechanics in elastic solids, International
  Journal of Fracture 79~(4) (1996) R63--R68.

\bibitem{balankin1996b}
A.~S. Balankin, The effect of fracture surface morphology on the crack
  mechanics in a brittle material, International Journal of Fracture 76 (1996)
  R63--R70.

\bibitem{balankin1997}
A.~S. Balankin, Physics of fracture and mechanics of self-affine cracks,
  Engineering Fracture Mechanics 57~(2-3) (1997) 135--203.

\bibitem{yavari2000}
A.~Yavari, K.~G. Hockett, S.~Sarkani, The fourth mode of fracture in fractal
  fracture mechanics, International Journal of Fracture 101~(4) (2000)
  365--384.

\bibitem{yavari2002}
A.~Yavari, S.~Sarkani, E.~T. Moyer, The mechanics of self-similar and
  self-affine fractal cracks, International Journal of Fracture 114~(1) (2002)
  1--27.

\bibitem{yavari2002b}
A.~Yavari, Generalization of barenblatt's cohesive fracture theory for fractal
  cracks, Fractals 10~(02) (2002) 189--198.

\bibitem{wnuk2003}
M.~P. Wnuk, A.~Yavari, On estimating stress intensity factors and modulus of
  cohesion for fractal cracks, Engineering Fracture Mechanics 70~(13) (2003)
  1659--1674.

\bibitem{wnuk2008}
M.~P. Wnuk, A.~Yavari, Discrete fractal fracture mechanics, Engineering
  Fracture Mechanics 75~(5) (2008) 1127--1142.

\bibitem{wnuk2009}
M.~P. Wnuk, A.~Yavari, A discrete cohesive model for fractal cracks,
  Engineering Fracture Mechanics 76~(4) (2009) 548--559.

\bibitem{yavari2010}
A.~Yavari, H.~Khezrzadeh, Estimating terminal velocity of rough cracks in the
  framework of discrete fractal fracture mechanics, Engineering Fracture
  Mechanics 77~(10) (2010) 1516--1526.

\bibitem{khezrzadeh2011}
H.~Khezrzadeh, M.~P. Wnuk, A.~Yavari, Influence of material ductility and crack
  surface roughness on fracture instability, Journal of Physics D: Applied
  Physics 44~(39) (2011) 395302.

\bibitem{carpinteri1994}
A.~Carpinteri, Fractal nature of material microstructure and size effects on
  apparent mechanical properties, Mechanics of materials 18~(2) (1994) 89--101.

\bibitem{mecholsky2020}
J.~Mecholsky~Jr, N.~Mecholsky, D.~DeLellis, Relationship of energy to geometry
  in brittle fracture, Journal of the European Ceramic Society 40~(13) (2020)
  4602--4604.

\bibitem{barenblatt1962}
G.~I. Barenblatt, The mathematical theory of equilibrium cracks in brittle
  fracture, Advances in applied mechanics 7 (1962) 55--129.

\bibitem{dugdale1960}
D.~S. Dugdale, Yielding of steel sheets containing slits, Journal of the
  Mechanics and Physics of Solids 8~(2) (1960) 100--104.

\bibitem{park2011}
K.~Park, G.~H. Paulino, Cohesive zone models: a critical review of
  traction-separation relationships across fracture surfaces, Applied Mechanics
  Reviews 64~(6) (2011) 1--20.

\bibitem{lurie2014}
S.~Lurie, P.~Belov, Gradient effects in fracture mechanics for nano-structured
  materials, Engineering Fracture Mechanics 130 (2014) 3--11.

\bibitem{vasil2016}
V.~Vasil’ev, S.~Lurie, New solution of the plane problem for an equilibrium
  crack, Mechanics of Solids 51~(5) (2016) 557--561.

\bibitem{vasil2020}
V.~Vasil’ev, S.~Lur’e, New method for studying the strength of brittle
  bodies with cracks, Russian Metallurgy (Metally) 2020 (2020) 291--297.

\bibitem{griffith1921}
A.~A. Griffith, Vi. the phenomena of rupture and flow in solids, Philosophical
  Transactions of the Royal Society of London. Series A, containing papers of a
  mathematical or physical character 221~(582-593) (1921) 163--198.

\bibitem{tvergaard1990}
V.~Tvergaard, Material failure by void growth to coalescence, Advances in
  Applied Mechanics 27~(1) (1990) 83--151.

\bibitem{pardoen2000}
T.~Pardoen, J.~Hutchinson, An extended model for void growth and coalescence,
  Journal of the Mechanics and Physics of Solids 48~(12) (2000) 2467--2512.

\bibitem{gurson1977}
A.~L. Gurson, Continuum theory of ductile rupture by void nucleation and
  growth: Part {I}. {Y}ield criteria and flow rules for porous ductile media,
  Journal of Engineering Materials and Technology 99~(1) (1977) 2--15.

\bibitem{tvergaard1984}
V.~Tvergaard, A.~Needleman, Analysis of the cup-cone fracture in a round
  tensile bar, Acta Metallurgica 32~(1) (1984) 157--169.

\bibitem{mcclintock1968}
F.~A. McClintock, A criterion for ductile fracture by the growth of holes,
  Journal of Applied Mechanics 35~(2) (1968) 363--371.

\bibitem{li2004}
S.~Li, G.~Wang, On damage theory of a cohesive medium, International Journal of
  Engineering Science 42~(8) (2004) 861--885.

\bibitem{green1968}
A.~E. Green, W.~Zerna, Theoretical Elasticity, Courier Corporation, 1992.

\bibitem{kassir1975}
M.~Kassir, G.~C. Sih, Three-dimensional crack problems: A new selection of
  crack solutions in three-dimensional elasticity, Noordhoff International
  Publishing 2 (1975).

\bibitem{chen1993}
W.~Chen, L.~Keer, Mixed-mode fatigue crack propagation of penny-shaped cracks,
  Journal of Engineering Materials and Technology 115~(4) (1993) 365--372.

\bibitem{orowan1954}
E.~Orowan, Energy criteria of fracture, Tech. rep., Massachusetts Inst of Tech
  Cambridge Dept of Mechanical Engineering (1954).

\bibitem{irwin1958}
G.~Irwin, Fracture in “handbuch der physik,” vol. v, Springer-Verlag
  Berlin, G{\"o}ttingen, Heidelberg (1958).

\bibitem{novozhilov1969}
V.~Novozhilov, On a necessary and sufficient criterion for brittle strength,
  Journal of Applied Mathematics and Mechanics 33~(2) (1969) 201--210.

\bibitem{pugno2004}
N.~M. Pugno†, R.~S. Ruoff‡, Quantized fracture mechanics, Philosophical
  Magazine 84~(27) (2004) 2829--2845.

\bibitem{gol1991}
R.~Gol'dshtein, A.~Mosolov, Cracks with a fractacl surface, Soviet Physics
  Doklady 36 (1991) 603.

\bibitem{mandelbrot1985}
B.~B. Mandelbrot, Self-affine fractals and fractal dimension, Physica Scripta
  32~(4) (1985) 257--260.

\bibitem{bouchaud1990}
E.~Bouchaud, G.~Lapasset, J.~Planes, Fractal dimension of fractured surfaces: a
  universal value, EPL (Europhysics Letters) 13~(1) (1990) 73.

\bibitem{maaloy1992}
K.~J. M{\aa}l{\o}y, A.~Hansen, E.~L. Hinrichsen, S.~Roux, Experimental
  measurements of the roughness of brittle cracks, Physical Review Letters
  68~(2) (1992) 213.

\bibitem{bouchaud1997}
E.~Bouchaud, Scaling properties of cracks, Journal of Physics: Condensed Matter
  9~(21) (1997) 4319--4344.

\bibitem{bouchaud2003}
E.~Bouchaud, The morphology of fracture surfaces: A tool for understanding
  crack propagation in complex materials, Surface Review and Letters 10~(05)
  (2003) 797--814.

\bibitem{ponson2006}
L.~Ponson, D.~Bonamy, H.~Auradou, G.~Mourot, S.~Morel, E.~Bouchaud, C.~Guillot,
  J.~P. Hulin, Anisotropic self-affine properties of experimental fracture
  surfaces, International Journal of Fracture 140~(1-4) (2006) 27--37.

\bibitem{quinn1999}
J.~B. Quinn, Extrapolation of fracture mirror and crack-branch sizes to large
  dimensions in biaxial strength tests of glass, Journal of the American
  Ceramic Society 82~(8) (1999) 2126--2132.

\bibitem{doquet2013}
V.~Doquet, N.~B. Ali, A.~Constantinescu, X.~Boutillon, Fracture of a
  borosilicate glass under triaxial tension, Mechanics of materials 57 (2013)
  15--29.

\bibitem{chopra2018}
S.~Chopra, K.~A. Deshmukh, A.~Deshmukh, D.~Peshwe, Inflorescence type
  morphology and mirror--mist--hackle pattern in tensile fractograph of
  mwcnt/pbt nano-composites, International Journal of Materials Research
  109~(6) (2018) 561--568.

\bibitem{bradt2011}
R.~C. Bradt, The fractography and crack patterns of broken glass, Journal of
  Failure Analysis and Prevention 11~(2) (2011) 79--96.

\bibitem{arakawa1991}
K.~Arakawa, K.~Takahashi, Relationships between fracture parameters and
  fracture surface roughness of brittle polymers, International Journal of
  Fracture 48~(2) (1991) 103--114.

\bibitem{weisstein2002}
E.~W. Weisstein, CRC Concise Encyclopedia of Mathematics, CRC press, 2002.

\bibitem{dormieux2006}
L.~Dormieux, D.~Kondo, F.~J. Ulm, Microporomechanics, John Wiley \& Sons, 2006.

\bibitem{trustrum1979}
K.~Trustrum, A.~D.~S. Jayatilaka, On estimating the weibull modulus for a
  brittle material, Journal of Materials Science 14~(5) (1979) 1080--1084.

\bibitem{matsuo1986}
Y.~Matsuo, K.~Kitakami, S.~Kimura, Flaw-size distribution of structural
  ceramics, International Journal of High Technology Ceramics 3~(2) (1987) 171.

\bibitem{matsuo1987}
Y.~Matsuo, K.~Kitakami, S.~Kimura, Crack size and strength distribution of
  structural ceramics after non-destructive inspection, Journal of Materials
  Science 22~(6) (1987) 2253--2256.

\bibitem{danzer2007}
R.~Danzer, P.~Supancic, J.~Pascual, T.~Lube, Fracture statistics of
  ceramics--weibull statistics and deviations from weibull statistics,
  Engineering Fracture Mechanics 74~(18) (2007) 2919--2932.

\bibitem{khezrzadeh2016}
H.~Khezrzadeh, Overall properties of particulate composites with fractal
  distribution of fibers, Mechanics of Materials 96 (2016) 1--11.

\bibitem{khezrzadeh2017}
H.~Khezrzadeh, A statistical micromechanical multiscale method for
  determination of the mechanical properties of composites with periodic
  microstructure, Composites Part B: Engineering 115 (2017) 138--143.

\bibitem{kfouri1979}
A.~Kfouri, Crack separation energy-rates for the {DBCS} model under biaxial
  modes of loading, Journal of the Mechanics and Physics of Solids 27~(2)
  (1979) 135--150.

\bibitem{mura1987}
T.~Mura, Micromechanics of defects in solids, International Journal of Fracture
  31 (1987) 233--242.

\bibitem{wnuk1990}
M.~P. Wnuk, Mathematical Modelling of Nonlinear Phenomena in Fracture
  Mechanics, Springer, 1990.

\bibitem{blal2012}
N.~Blal, L.~Daridon, Y.~Monerie, S.~Pagano, Artificial compliance inherent to
  the intrinsic cohesive zone models: criteria and application to planar
  meshes, International Journal of Fracture 178~(1-2) (2012) 71--83.

\end{thebibliography}

\end{document}